\documentclass[11pt]{article}

\usepackage[long,no.tricks]{optional}

\usepackage[a4paper,top=33mm,bottom=33mm,left=24mm,right=24mm]{geometry}
\usepackage[utf8x]{inputenc}
\usepackage[english]{babel}
\usepackage{etex,musixtex}
\usepackage{algorithmic}
\usepackage{amsmath,amssymb,amsthm}
\usepackage{graphicx}
\usepackage{natbib}
\usepackage{txfonts}

\usepackage[pdftex,unicode,pdfborder={0 0 0},colorlinks=true,citecolor=blue,linkcolor=blue,urlcolor=blue]{hyperref}
\usepackage[all]{hypcap}
\usepackage[labelfont=bf,labelsep=colon]{caption}

\opt{ps.tricks}{
\usepackage{pstricks,pstricks-add}
\usepackage{auto-pst-pdf}
\def\dpi{6.2831853071795862}

\newcommand{\sinusoid}[2][1.2]{\mbox{}\\[-12pt]
  \begin{pspicture}(0,-#1)(4.2,#1)
  \psplot[plotpoints=1000,algebraic=true,linewidth=1pt]{0}{4}{#2}
  \psaxes[labels=none,ticks=none]{->}(0,0)(-0.2,-#1)(4.2,#1)
  \end{pspicture}}
}

\theoremstyle{definition}
\newtheorem{defn}{Definition}
\newtheorem{exmp}[defn]{Example}
\newtheorem{rem}[defn]{Remark}

\newcommand{\secref}[1]{Section~\ref{#1}}
\newcommand{\figref}[1]{Figure~\ref{#1}}
\newcommand{\tabref}[1]{Table~\ref{#1}}
\newcommand{\defref}[1]{Definition~\ref{#1}}
\newcommand{\exaref}[1]{Example~\ref{#1}}
\newcommand{\remref}[1]{Remark~\ref{#1}}

\newcommand{\cor}{\ensuremath{\hat{=}}}
\newcommand{\fix}[1]{\lfloor#1\rfloor}
\newcommand{\lcm}{\ensuremath{\operatorname{lcm}}}

\newcommand{\harm}[1]{\ensuremath{\{#1\}}}
\newcommand{\clas}[1]{\ensuremath{[#1]}}
\newcommand{\semi}[1]{\ensuremath{\{#1\}}}

\renewcommand{\H}{\ensuremath{\mathcal{H}}}
\newcommand{\T}{\ensuremath{\mathcal{T}}}

\newenvironment{keywords}{\medskip\noindent\textbf{Keywords:}}{}

\begin{document}

\hypersetup{pdfauthor={Frieder Stolzenburg},pdftitle={Harmony Perception by Periodicity Detection}}
\pdfbookmark[0]{Harmony Perception by Periodicity Detection}{toc}

\title{\bf Harmony Perception by Periodicity Detection\thanks{Original paper
	appeared in \emph{Journal of Mathematics and Music}, 9(3):215--238, 2015.
	DOI: \href{http://dx.doi.org/10.1080/17459737.2015.1033024}{10.1080/17459737.2015.1033024}}}
\author{Frieder Stolzenburg\thanks{E-mail:
	\href{mailto:fstolzenburg@hs-harz.de}{fstolzenburg@hs-harz.de}}\\[6pt] Harz University of Applied Sciences,
	Automation \& Computer Sciences Department,\\ Friedrichstr. 57-59,
	38855~Wernigerode, GERMANY}
\date{}
\maketitle

\begin{abstract}
The perception of consonance/dissonance of musical harmonies is strongly
correlated with their periodicity. This is shown in this article by consistently
applying recent results from psychophysics and neuroacoustics, namely that the
just noticeable difference between pitches for humans is about 1\% for the
musically important low frequency range and that periodicities of complex chords
can be detected in the human brain. Based thereon, the concepts of relative and
logarithmic periodicity with smoothing are introduced as powerful measures of harmoniousness.
The presented results correlate significantly with empirical investigations on the
perception of chords. Even for scales, plausible results are obtained. For
example, all classical church modes appear in the front ranks of all
theoretically possible seven-tone scales.

\end{abstract}

\section{Introduction}\label{sec:intro}

Music perception and composition seem to be influenced not only by convention or
culture, manifested by musical styles or composers, but also by the
neuroacoustics and psychophysics of tone perception \citep{Lan97,Roe08}. While
studying the process of musical creativity including harmony perception, several
questions may arise, such as: What are the underlying principles of music
perception? How can the perceived consonance/dissonance of chords and scales be
explained?

\begin{figure*}
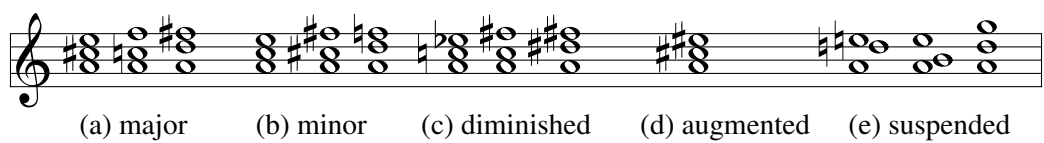

\centering
  \begin{music}
    \instrumentnumber{1}\nobarnumbers
      \startextract
	\Notes\zsong{(a) major}\zw{'a^ce}\enotes
	  \Notes\zw{'a=cf}\enotes
	  \NOTes\zw{'ad^f}\enotes
	\NOtes\zsong{(b) minor}\zw{'ace}\enotes
	  \Notes\zw{'ac^f}\loff{\sh c}\enotes
	  \NOTes\zw{'ad=f}\enotes
	\Notes\zsong{\hspace*{-10pt}(c) diminished}\loff{\na j}\zw{'ac_e}\enotes
	  \NOtes\zw{'ac^f}\enotes
	  \NOTes\zw{'ad^f}\loff{\sh d}\enotes
	\Notes\zsong{(d) augmented}\enotes
	  \NOTes\loff{\sh j}\zw{'ac^e}\enotes
	  \NOTes\enotes
	\NOtes\zsong{(e) suspended}\loff{\na k}\zw{'a=e}\rw{d}\enotes
	  \NOtes\zw{'ae}\rw{b}\enotes
	  \NOtes\zw{'adg}\enotes
      \endextract
  \end{music}
\caption{Common triads and their inversions. \figref{fig:triad}(a)--(d) show all
triads that can be built from major and minor thirds, i.e., the distance between
successive pairs of tones are three or four semitones, including their inversions
which are obtained by transposing the currently lowest tone by an octave, always
with A4 as the lowest tone here. \figref{fig:triad}(e) shows the suspended chord,
built from perfect fourths (five semitones apart). Its last inversion reveals
this.}\label{fig:triad}
\end{figure*}

Numerous approaches tackle these questions, studying the consonance/dissonance
of dyads and triads \citep{KK69,HK78,HK79,CF06,JKL12}.\opt{long}{ For instance,
the major triad (\figref{fig:triad}(a)) is often associated with emotional terms
like \emph{pleasant}, \emph{strong}, or \emph{bright}, and, in contrast to this,
the minor triad (\figref{fig:triad}(b)) with terms like \emph{sad}, \emph{weak},
or \emph{dark}.} Empirical studies reveal
a clear preference ordering on the perceived consonance/dissonance of common
triads in Western music (see \figref{fig:triad}), e.g. major $\prec$ minor
\citep{Rob86,JKL12}.

Early mathematical models relate musical intervals, i.e. the distance between
two pitches, to simple fractions by applying the fact that, in physical terms,
an interval is the ratio between two sonic frequencies. This helps to understand
that human subjects rate harmonies, e.g. major and minor triads, differently
with respect to their consonance.\opt{long}{ But since most common triads
(cf.~\figref{fig:triad}) throughout are built from thirds, thirds do not provide
a direct explanation of the preference ordering.} Newer explanations are based
on the notion of dissonance, roughness, instability, or tension
\citep{Hel63,HK78,HK79,TSS82,Par89,Set05,CF06,Coo09b,Coo12}. They correlate
better with empirical results on harmony perception, but still can be improved.

\subsection{Aims}

A theory of harmony perception should be as simple and explanatory as possible.
This means on the one hand, that it does not make too many assumptions, e.g. on
implicit or explicit knowledge about diatonic major scales from Western music
theory, or use many mathematical parameters determining dissonance or roughness
curves. On the other hand, it should be applicable to musical harmonies in a
broad sense: Tones of a harmony may sound simultaneously as in chords, or
consecutively and hence in context as in melodies or scales. A harmony can thus
simply be identified by a set of tones forming the respective interval,
chord, or scale. This article aims at and presents a fully computational (and
hence falsifiable) model for musical harmoniousness with high explanatory power
for harmonies in this general, abstract sense. 

A theory of harmony perception should furthermore also consider and incorporate
the results from neuroacoustics and psychophysics on auditory processing of
musical tone sensations in the ear and the brain. The frequency analysis in the
inner ear can be compared with that of a filter bank with many parallel channels.
Periodicities of
complex chords can be detected in the human brain, and information concerning
stimulus periodicities is still present in short-term memory
\citep{Lan97,LS+09}.

Finally, the correlation between the consonance/dissonance
predicted by the theory and the perceived consonance/dissonance from empirical
studies should be the highest possible. In order to make the theory definitely
non-circular, we therefore want to establish the relation between
musical harmonies, i.e. complex vibrations, and their perceived
consonance/dissonance. \opt{long}{Several empirical experiments on harmony
perception have been conducted \citep{Mal18,Rob86,Kru90,JKL12,TT13}. For this,
participants are asked to listen to and rate musical harmonies on an ordinal
scale with respect to their consonance/dissonance. We will use the results of
these experiments, which take intervals, chords, or scales as stimuli, to
evaluate theories on harmony perception (in~\secref{sec:eval}).}

\subsection{Main contribution}

This article applies recent results from psychophysics and neuroacoustics
consistently, obtaining a fully computational theory of consonance/dissonance
perception. We focus on periodicity detection, which can be identified as a
fundamental mechanism to music perception, and exploit in addition the fact that
the just noticeable difference between pitches for humans is about 1\% for the
musically important low frequency range \citep{ZFS57,Roe08}. This percentage is
the only parameter of the present approach. The concept of \emph{periodicity pitch}
has been studied for intervals many times in the literature \citep[cf.][and
references therein]{Roe08}. The idea in this article is to transfer this concept
to chords and also scales by considering relative periodicity, that is
the approximated ratio of the period length of the chord (its periodicity pitch)
relative to the period length of its lowest tone component (without necessarily
applying octave equivalence), called \emph{harmonicity} $h$ in \citet{ICMPC.Sto09}.
In this article we will use the term \emph{relative periodicity} rather than
harmonicity.

The hypothesis in this article is that the perceived consonance of a musical
harmony decreases as the relative periodicity $h$ increases. The corresponding
analysis presented here (in \secref{sec:eval}) confirms that it does not matter much
whether tones are presented consecutively (and hence in context) as in scales or
simultaneously as in chords. Periodicity detection seems to be an
important mechanism for the perception of all kinds of musical harmony: chords,
scales, and probably also chord progressions. Listeners always prefer simpler,
i.e. shorter periodic patterns. The predictions of the presented periodicity-based
method with smoothing obtained for dyads and common triads on the one hand and diatonic
scales, i.e. the classical church modes, on the other hand all show highest
correlation with the empirical results \citep{SHP03,JKL12,TT13}, not only with
respect to the ranks, but also with the ordinal values of the empirical ratings
of musical consonance. For the latter, we consider \emph{logarithmic periodicity}, i.e.
$\log_2(h)$. As we will see, this logarithmic measure can plausibly be motivated
by the concrete topological organisation of the periodicity coding in the brain
(cf.~\secref{sec:neuronal}).

\subsection{Overview of the rest of the article}

The organisation of the article is straightforward. After this introductory
section (\secref{sec:intro}), we briefly discuss existing theories on harmony
perception (\secref{sec:theo})\opt{long}{, which often make use of the notions
consonance and dissonance}. In particular, we highlight the psychophysical basis
of harmony perception by reviewing recent results from neuroacoustics on
periodicity detection in the brain \citep[e.g.][]{Lan97}. Then, we define
relative and logarithmic periodicity with smoothing in detail (\secref{sec:harm}). Applying
these measures of harmoniousness to common musical chords and also
scales shows very high correlation with empirical results (\secref{sec:eval}). We
compare these results with those of other models of harmony perception in our
evaluation. Finally, we draw some conclusions (\secref{sec:sum}).

\section{Theories of harmony perception}\label{sec:theo}

Since ancient times, the problem of explaining musical harmony perception
attracted a lot of interest. In what follows, we discuss some of them briefly. But
since there are numerous approaches addressing the problem, the following list
is by no means complete. We mainly concentrate on theories with a psychophysical
or neuroacoustical background and on those that provide a mathematical, i.e. computational measure
of consonance/dissonance that we can use in our evaluation of the different
approaches (in~\secref{sec:eval}).

\subsection{Overtones}

A harmonic complex tone may consist of arbitrary sinusoidal tones, called
\emph{partials}, with harmonically related frequencies. The frequency of the
$n$th (harmonic) partial is $f_n = n \cdot f$, where $n \ge 1$ and $f=f_1$ is
the frequency of the fundamental tone. The tones produced by real instruments,
such as strings, tubes, or the human voice, have harmonic or other
\emph{overtones}, where the $n$th overtone corresponds to the $(n\!+\!1)$st
partial.\opt{long}{

}Overtones\opt{long}{ (cf.~\figref{fig:ovt})} may explain the origin of
the major triad and hence its high perceived consonance. The major triad appears
early in the sequence, namely partials $4$, $5$, $6$ (root position) and -- even
earlier -- $3$, $4$, $5$ (second inversion). But this does not hold for the minor
chord. The partials $6$, $7$, $9$ form a minor chord, which is out of tune
however, since the frequency ratio $7/6$ differs from the ratio $6/5$, which is
usually assumed for the minor third. The first minor chord in tune appears
only at the partials $10$, $12$, $15$.\opt{long}{ In contrast to the major
triad, the partials forming the minor triad are not adjacent, and their ground
tones (B4 and G$\sharp$5, respectively) are not octave equivalent (i.e., they do
not have the same basic name) with the fundamental tone E2 of the overtone
series.

Furthermore, a diminished triad is given by the partials $5$, $6$, $7$.
Therefore, it appears previous to the minor triad in the overtone series, which
is inconsistent with the empirical results. Nevertheless, the so obtained
diminished triad is also out of tune. The intervals correspond to different
frequency ratios, namely $6/5$ and $7/6$, although the diminished triad is built
from two (equal) minor thirds.

\begin{figure}
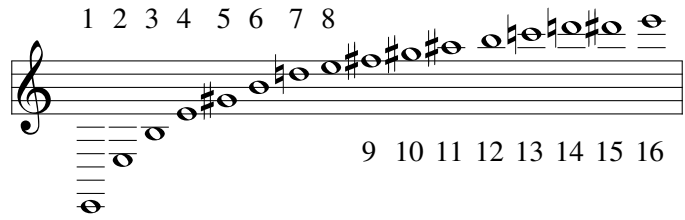

\centering
  \begin{music}
    \instrumentnumber{1}
    \startextract
	\notes\zcharnote{13}{1}\wh{E}\enotes
	\notes\zcharnote{13}{2}\wh{'E}\enotes
	\notes\zcharnote{13}{3}\wh{b}\enotes
	\notesp\zcharnote{13}{4}\wh{e}\enotes
	\notes\zcharnote{13}{5}\wh{^g}\enotes
	\notesp\zcharnote{13}{6}\wh{'b}\enotes
	\notes\zcharnote{13}{7}\wh{'=d}\enotes
	\notesp\zcharnote{13}{8}\wh{'e}\enotes
	\notesp\csong{9}\wh{'^f}\enotes
	\notesp\csong{10}\wh{'^g}\enotes
	\notesp\csong{11}\wh{''^a}\enotes
	\notesp\csong{12}\wh{''b}\enotes
	\notesp\csong{13}\wh{''=c}\enotes
	\notesp\csong{14}\wh{''=d}\enotes
	\notesp\csong{15}\wh{''^d}\enotes
	\notesp\csong{16}\wh{''e}\enotes
    \endextract
  \end{music}
\caption{Harmonic partials (approximated) with respect to the fundamental tone E2.}\label{fig:ovt}
\end{figure}}

\subsection{Frequency ratios}\label{sec:ratio}

Already Pythagoras studied harmony perception, relating music and mathematics
with each other. He created a tuning system that minimised roughness for
intervals of harmonic complexes and expressed musical intervals as simple
fractions, i.e. with small numerators and denominators. Also
\citet{Eul39} \citep[see also][]{Bai97} spent some time with the problem and
proposed the so-called \emph{gradus suavitatis} (degree of softness) as a
measure of harmoniousness, considering the least common multiple (\lcm) of
denominators and numerators of the frequency ratios \harm{a_1/b_1, \ldots,
a_k/b_k} (where all fractions $a_i/b_i$ are in lowest terms) with respect to the
lowest tone in the given harmony. If $p_1^{m_1} \cdot\ldots\cdot p_\ell^{m_\ell}$
is the prime factorisation of $n = \lcm(a_1,\ldots,a_k) \cdot \lcm(b_1, \ldots,
b_k)$, then the gradus suavitatis is defined by $\Gamma(n) = 1+\sum_{i=1}^\ell
m_i\cdot(p_i-1)$. For instance, for the major second interval consisting of the
frequency ratios \harm{1/1,9/8} (including the root), we have $\lcm(1,9)
\cdot \lcm(1,8) = 72 = 2^3 \cdot 3^2$ and hence $\Gamma(72) = 1+3\cdot 1+2\cdot 2=8$
as the value of the gradus suavitatis.

Following somewhat the lines of \citet{Eul39}, \citet{ICMPC.Sto12b} considers,
besides periodicity, the complexity of the product $\lcm(a_1,\ldots,a_k) \cdot
\lcm(b_1,\ldots,b_k)$ with respect to prime factorisation. More precisely, it is
the number of not necessarily distinct prime factors $\Omega$
\citep[cf.][p.\,354]{HW79}, defined for natural numbers as follows: $\Omega(1) =
0$, $\Omega(n)$ = 1, if $n$ is a prime number, and $\Omega(n) =
\Omega(n_1)+\Omega(n_2)$ if $n=n_1 \cdot n_2$. For example, for the major second
interval (see above), we have $\Omega(72) = \Omega(2^3 \cdot 3^2) = 3+2 =
5$.\opt{long}{ The function shares properties with the logarithm function (cf.
last part of definition for composite numbers).} The rationale
behind this measure is to count the maximal number of times that the whole
periodic structure of the given chord can be decomposed in time intervals of
equal length. \citet{Bre05} proposes $(a_1 \cdot\ldots\cdot a_k \cdot b_1
\cdot\ldots\cdot b_k)^{1/(2k)}$ as a measure of harmoniousness, i.e. the geometric
average determined from the numerators and denominators of the frequency ratios
of all involved intervals. Both these measures yield reasonable harmoniousness
values in many cases.

\opt{long}{\citet{HB05,HB11} investigate convexity of scales, by visualising
them on the Euler lattice, in which each point represents an integer power of
prime factors, where also negative exponents are allowed. Because of octave
equivalence (which is adopted in that approach), the prime $2$ is omitted. For
example, the frequency ratio $5/3$ (major sixth) can be written as
$3^{-1}\cdot5^1$ and thus be visualised as point $(-1,1)$. \citet{HB05,HB11}
consider periodicity blocks according to \citet{Fok69} and observe that almost
all traditional scales form (star-)convex subsets in this space. However, the
star-convexity property does not allow to rank harmonies, as done in this
article, at least not in a direct manner, because it is a multi-dimensional
measure.}

\subsection{Dissonance, roughness, and instability}\label{sec:diss}

\citet{Hel63} explains the degree of \emph{consonance} in terms of coincidence
and proximity of overtones and difference tones. For instance, for the minor
second (frequency ratio $16/15$), only very high, low-energy overtones coincide,
so it is weakly consonant. For the perfect fifth (frequency ratio $3/2$), all
its most powerful overtones coincide, and only very weak ones are close enough
to beat. The perfect fifth is therefore strongly consonant and only weakly
dissonant. The theory of \citet{Hel63} has already been criticised convincingly
by contemporaries such as R.\,H.~Lotze, C.~Stumpf, or E.~Mach, but it has
survived with some modifications until today \citep[e.g.][]{PL65}. Newer
explanations are based on the notion of \emph{dissonance} or \emph{roughness}
\citep{KK69,HK78,HK79,Par89,Set05}. Other important explanations refer to tonal
fusion \citep{Stu90} or combination tones \citep{Kru04}.

\opt{long}{In general, dissonance is understood as the opposite to consonance,
meaning how well tones sound together. If two sine waves sound together, typical
perceptions include pleasant beating (when the frequency difference is small),
roughness (when the difference grows larger), and separation into two tones
(when the frequency difference increases even further) \citep[Figure~3.7]{Set05}.
Based on these observations, several mathematical functions for dissonance or
roughness curves are proposed in the literature. \figref{fig:diss} shows one
possibility according to \citet{HK78,HK79} with two parameters $a$ and $b$,
where $y$ denotes the sensory dissonance and $x$ is the relative deviation
between the two frequencies. In order to find out the dissonance among several
tones, usually their overtone spectra are also taken into account, summing up
the single dissonance values.

\begin{figure}
\centering
\opt{no.tricks}{\includegraphics[page=1]{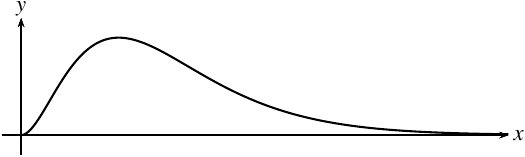}}
\opt{ps.tricks}{
\def\eul{2.7182818284590455}
\psset{unit=1.65}
    \begin{pspicture}(-0.2,-0.2)(5.1,1.3)
	\psplot[plotpoints=1000,algebraic=true,linewidth=1pt]{0.0}{5.0}{(x*exp(\eul,(1-x)))^2}
	\psaxes[labels=none,ticks=none]{->}(0,0)(-0.2,-0.2)(5.0,1.2)
	\rput{0}(5.1,0){$x$}
	\rput{0}(0,1.3){$y$}
    \end{pspicture}
}
\caption{Dissonance or roughness curve with $y = \left(x/a \cdot
\exp\left(1-x/a\right)\right)^b$ and $x = |f_1/f_0-1|$ (absolute value of
relative frequency deviation) following the lines of \citet{HK78,HK79}. $a$ is
the interval for maximum roughness, and $b$ is an index set to $2$ to yield the
standard curve.}\label{fig:diss}
\end{figure}
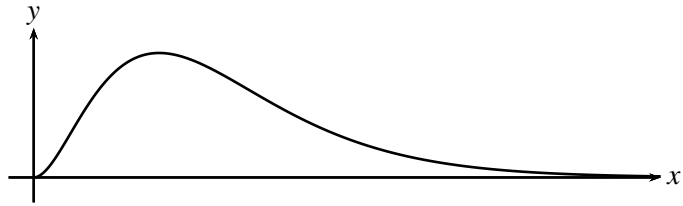}

Although these approaches correlate better with the empirical results on harmony
perception, they do not explain the low perceived consonance of the diminished
or the augmented triad, which are built from two minor or major thirds,
respectively. Therefore, \citet{CF06} emphasise that harmony is
more than the summation of interval dissonance among tones and their upper
partials, adopting the argument from psychology that neighbouring intervals of
equal size are unstable and produce a sense of tonal \emph{instability} or
\emph{tension}, that is resolved by pitch changes leading to unequal intervals
\citep[see also][]{Coo12}.\opt{long}{ Lowering any tone in an augmented triad by
one semitone leads to a major triad and raising to a minor triad. Because of
this, \citet{CF06} assume sound symbolism, where the major triad is associated
with social strength and the minor triad with social weakness.

Since overtone spectra vary largely among different instruments, it is difficult
to determine the number and amplitudes of overtones. This uncertainty makes the
calculation of dissonance as sketched above a bit vague. Maybe because of this,
works based on dissonance and related notions focus on the consonance of dyads
and triads, whereas the analysis of more complex chords or even scales is less often
studied.

\subsection{Human speech and musical categories}

\citet{GP09}\opt{long}{ \citep[see also][]{BP15}} consider a more biological rationale, investigating the
relationship between musical modes and vocalisation. They examine the hypothesis
that major and minor scales elicit different affective reactions because their
spectra are similar to the spectra of voiced speech segments. Their results reveal
that spectra of intervals in major scales are more similar to the ones found in
excited speech, whereas spectra of particular intervals in minor scales are more
similar to the ones of subdued speech \citep{BG+10}. The observation, that the
statistical structure of human speech correlates with common musical categories,
can also be applied to consonance rankings of dyads, yielding plausible results
\citep{SHP03}. As a measure for comparing harmonies, the mean \emph{percentage
similarity} of the respective spectra is considered in this context. For an
interval with frequency ratio $a/b$, it is defined as $(a+b-1)/(a \cdot b)$
\citep[p.~2]{GP09}.}

\subsection{Periodicity-based approaches}\label{sec:period}

The approaches discussed so far essentially take the frequency spectrum of a
sound as their starting point.
Clearly, analysing the frequency spectrum is closely related to analysing the
time domain (with respect to periodicity). Fourier transformation allows us to translate between
both mathematically. However, subjective pitch detection, i.e. the capability of
the auditory system to identify the repetition rate of a complex tone, only
works for the lower but musically important frequency range
up to about 1500\,Hz~\citep{Plo67}. A \emph{missing fundamental} tone
can be assigned to each interval. For this, the frequency components of the
given interval or chord are interpreted as (harmonic) overtones of a common
fundamental frequency. Note that the tone with the fundamental frequency, called the
\emph{periodicity pitch} of the interval, is not always present as an original tone
component (and hence missing).\opt{long}{ Therefore, periodicity pitch and tone
pitch represent distinct dimensions of harmony perception.}

\citet{Sch38,Sch40} introduces the notion of \emph{residue} for periodicity pitch.
Based on these works, several periodicity-based approaches have
been proposed \citep{Lic51,Lic62,BC61,TSS82,Bea01}. \citet{Lic62} discusses
possible neuronal mechanisms for periodicity pitch, including autocorrelation
and comb-filtering, which has been proved correct by recent results from
neuroacoustics (see \secref{sec:neuronal}). \citet{BC61} explain musical
phenomena as relationships between long patterns of waves, using so-called
\emph{frequency discs}. \citet{Hof04,Hof08} provides a more computational model,
which determines the most probable missing fundamental tone in the so-called
\emph{sonance factor}. Relative periodicity $h$ presented in this article (in
\secref{sec:harm}) is also based on periodicity detection. It is a fully
computational model of consonance/dissonance, applicable to harmonies in the
broad sense, which shows high correlation with empirical ratings.

\subsection{Neuronal models}\label{sec:neuronal}

Since musical harmony seems to be a phenomenon
present in almost all human cultures, harmony perception must be somehow closely
connected with the auditory processing of musical tone sensations in the ear and
in the brain. That periodicity can be detected in the brain has
been well known for years.\opt{long}{ For example, two pure tones forming a
mistuned octave cause so-called \emph{second-order beats}, although no exact
octave is present \citep{Plo67,Roe08}.} But only recently has neuroacoustics
found the mechanism for being able to perceive periodicity, which lays the basis
for the notion of relative periodicity $h$ introduced in this article:
\citet{Lan97} successfully conducted experiments on periodicity detection with
animals and humans.\opt{long}{ The resulting neuronal model \citep{Lan97\opt{long}{,Lan15}}
for the analysis of periodic signals is sketched in \figref{fig:langner}. It
contains several stages, that we will explain next.

\begin{figure}
\centering
\includegraphics[width=0.45\textwidth]{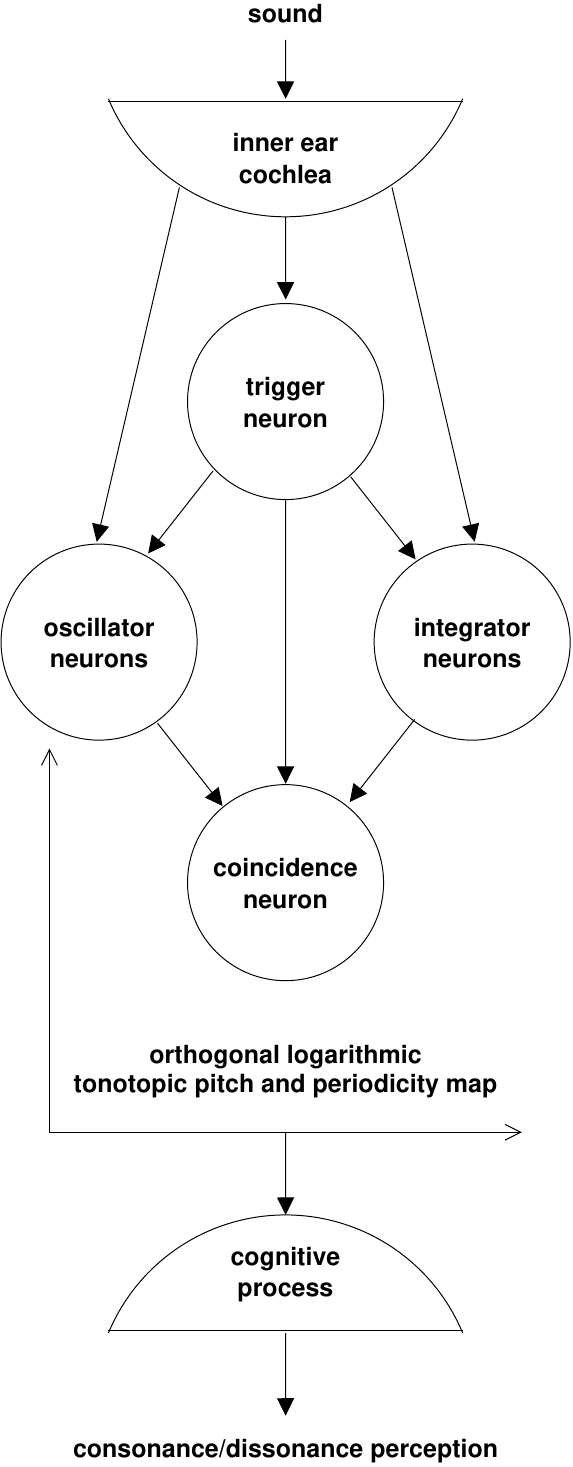} 
\caption{Sketch of the neuronal model for the analysis of periodic signals
\citep[cf.][Figure~3]{Lan97}. The arrows roughly denote the workflow of the
auditory processing.}\label{fig:langner}
\end{figure}

First, so-called \emph{trigger neurons} in the ventral cochlear nucleus, well
known as octopus cells in anatomical terms, transfer signals without significant
delay. The period of the original signal and the neuronal activity correspond,
but the neuronal activity typically has the form of spike trains, i.e. a series
of discrete action potentials \citep{LS_88,Car99,TC+01}. The maximal amplitude
of these spikes is limited. This means, that some information on the waveform of
the original signal concerning the overtone spectrum and also the amplitudes,
which contribute to the timbre and loudness of the sound, is lost and hence in
consequence the original signal is highly distorted by this.

Second, there are \emph{oscillator neurons}, which are chopper neurons or
stellate cells, with intrinsic oscillation showing regularly timed discharges in
response to stimuli, not corresponding to the temporal structure of the external
signal. The oscillation intervals can be characterised as integer multiples $n
\cdot T$ of a base period of $T=0.4$\,ms with $n \ge 2$ for endothermic, i.e.
warm-blooded animals \citep{Lan83,SL97,BL06}. The external signal is
synchronised with that of the oscillator neurons, which limits signal
resolution. Hence, frequencies and also intervals can only be distinguished up
to a certain precision.

Third, in the dorsal cochlear nucleus, periodic signals are transferred with
(different) delays. Onset latencies of so-called \emph{integrator neurons}
(fusiform cells, i.e. type IV cells, and giant cells in anatomical terms) up to
120\,ms have been observed \citep{LS_88}. While oscillator neurons respond with
almost no delay, integrator neurons respond with a certain amount of delay. Both
groups of neurons are triggered and synchronised by trigger neurons (on-cells).
When the delay corresponds to the signal period, the delayed response and the
non-delayed response to the next modulation wave coincide. All neurons respond
also to the harmonic overtones of their characteristic frequency \citep{Lan97}.
Therefore, the missing fundamental tone and hence periodicity pitch can be
detected in the brain. In the auditory midbrain (inferior colliculus),
\emph{coincidence neurons} respond best whenever the delay is compensated by the
signal period.

The latter provides the basis for an autocorrelation mechanism by
comb-filtering, which includes \emph{phase locking} \citep{Lan83,MH91,LS+09}.
This means, phase differences among different signals can be neglected.
Furthermore, runtime differences between different spike trains in the auditory
system are nullified, thus facilitating the highest possible coincidence rate
between two correlated spike trains. The neurons in the midbrain inferior
colliculus are capable of phase locking to stimulus periodicities up to
1000\,Hz, as \citet{LS_88} found out by experiments with cats. There seems to be
evidence that periodicity detection is possible for significantly higher
frequencies in some cases \citep[cf.~\secref{sec:period}]{Lan97}.} Pitch and timbre
(i.e. frequency and periodicity) are mapped temporally and also spatially and
orthogonally to each other in the auditory midbrain and auditory cortex as a
result of a combined frequency-time analysis that is some kind of
autocorrelation or comb-filtering. The neuronal periodicity maps in the inner
cortex (of cat and gerbil) and cortex (of gerbil and human) are organised along
logarithmic axes in accordance with the frequency tuning of the respective
neurons \citep{LS_88,SL97,LS+97}. In total, about $8$ octaves can be represented
in each dimension.

\citet{LS+09}\opt{long}{ \citep[see also][]{LS+15}} discovered that the phase-locking activity to the temporal
envelope is more accurate (i.e. sharper) in musicians than non-musicians,
reporting experiments with two intervals: The consonant interval (major sixth,
tones with 99 and 166\,Hz, approximate frequency ratio $5/\underline{3}$)
shows the highest response in the brainstem at about
30.25\,ms\,\cor\,99/\underline{3}\,Hz, and the dissonant interval (minor
seventh, 93 and 166\,Hz, frequency ratio 9/\underline{5}) at about
54.1\,ms\,\cor\,93/\underline{5}\,Hz. All this fits very well with
periodicity-based approaches to harmony perception: The denominator
of the approximate frequency ratio (underlined above) is the relative period
length with respect to the frequency of the lowest tone, as we shall see later.
In summary, the periodicity of a complex tone sensation can be detected by the
auditory system in the brain. In addition, it seems possible that the auditory
system alters the spectrum of intervals making them sound more consonant, i.e.
obtaining simpler frequency ratios.

\opt{long}{\subsection{Autocorrelation and tonal fusion}}

All these results, which are also consistent with results from fMRI~experiments
on melody processing in the auditory pathway \citep{PU+02}, indicate that
periodicity detection in the brain may be an important mechanism during music
perception. Therefore, several approaches to harmony perception
\citep{Ebe07,Ebe08,ICMPC.Fol12} directly refer to \citet{Lan97}.
\citet{Ebe07,Ebe08}\opt{long}{ \citep[see also][]{Ebe14}} develops a neuronal model of pulse chains in the auditory
systems testing different pulse forms. Applying autocorrelation functions, he
derives the so-called generalised coincidence function by computing the energy
of the autocorrelation function. It provides a periodicity measure for arbitrary
stimuli. Intervals do not need to be given by ratios of positive integers,
arbitrary vibration ratios can be handled. It turns out that this mathematical
model defines a measure that is in line with the degree of \emph{tonal fusion}
as described by \citet{Stu83,Stu90,Stu98}.

\opt{long}{\citet{Stu83,Stu90} assumes that listeners recognise whether one or
more tones are present and how similar or different they are. The perception of
musical harmony arises then after tonal fusion, where the sensation of several
tones results into a single impression. From introspection and extensive hearing
experiments, in particular with musically not trained persons, he deduces a
system of rules for the degree of tonal fusion and illustrates it in a curve
which shows the degree of fusion for all intervals over a range of one octave
\citep[Section~19, p.\,176]{Stu90}. According to \citet[Section~8]{Stu98}, the
fusion of two tones is not changed when further tones are added. Nonetheless,
musical harmony should be considered holistically. It is more than the summation
of interval consonance/dissonance. Already \cite{CF06} applied this principle of
gestalt psychology to triads (cf.~\secref{sec:diss}). In this article, this idea
is developed even further and applied to arbitrarily complex harmonies. Relative
and logarithmic periodicity, which we introduce in detail in \secref{sec:harm},
always refer to the harmony as a whole.}

\citet{Car99} reviews
neurophysiological evidence for interspike interval-based representations for
pitch and timbre in the auditory nerve and cochlear nucleus \citep[see
also][]{TC+01}. Timings of discharges in auditory nerve fibres reflect the time
structure of acoustic waveforms, such that the interspike intervals (i.e. the
period lengths) that are produced convey information concerning stimulus
periodicities that are still present in short-term memory. Common to these
approaches is that they focus on the autocorrelation function within one period
of the given harmony, considering the interspike intervals between both
successive or non-successive spikes in a discharge pattern. This leads to
histograms, called \emph{autocorrelograms}, which show high peaks for periods
corresponding to the pitch. This procedure works well for dyads, but for triads
and more complex chords the correlation with empirical ratings is relatively
low, which has already been noticed by~\citet[Section~2.5.3]{Ebe07}.

\opt{long}{\citet{MM+13} conclude from experiments with common dyads and
triads that the perception of consonance/dissonance involves cognitive
processes. According to the cognitive incongruence model of dissonance,
proposed by \citet{MM+13}, consonance/dissonance is learned to a certain extent
and is based on enculturation, i.e. the natural process of learning a particular
culture. Nevertheless, also in this model, periodicity plays a prominent role
\citep[Figure~10]{MM+13}. So, consistent with the results by \citet{LS+09}, a
general tendency of musically trained persons toward higher dissonance ratings
is observed, as musicians become better able to use periodicity-based pitch
mechanisms.}

\subsection{Cognitive\opt{long}{ and further} theories}

Cognitive science is the interdisciplinary scientific study of the mind and its
processes. It includes research on intelligence and behaviour, especially
focusing on how information is represented, processed, and transformed. There
exists also a series of approaches on harmony perception following this
paradigm, in particular for more complex chords and scales. For instance,
\citet{JKL12} propose a cognitive \emph{dual-process theory}, employing three
basic principles for ranking triads, namely (in order of  decreasing priority)
the use of diatonic scales, the central role of the major triad, and the
construction of chords out of thirds. These three principles of Western tonal
music predict a trend in increasing dissonance. Within each level of dissonance,
roughness according to \citet{HK78,HK79} can then predict a detailed rank order. 

\citet{TT13} investigate the emotional connotations of diatonic
modes. In the respective experiments, participants hear pairs of melodies,
presented in different diatonic modes, and have to judge which of the two
melodies sounds happier. The resulting sequence Ionian $\prec$ Mixolydian $\prec$ Lydian
$\prec$ Dorian $\prec$ Aeolian $\prec$ Phrygian (in descending order of preference) is
then explained by several principles: \emph{familiarity}, e.g. the major mode
(Ionian) is the most common mode in classical and popular music, and
\emph{sharpness}, i.e. the position of the scale relative to the tonic on the
circle of fifths. Both these cognitive theories show high correlation with the
empirical results with respect to their application domain, i.e. chords in
Western music and diatonic scales, respectively (cf.~\secref{sec:eval}). 

\opt{long}{There are other, more or less purely mathematical explanations for
the origin of chords and scales, e.g. by group theory \citep{Bal80,CC89},
ignoring however the sensory psychophysical basis. Therefore, we will not
consider these approaches further here.}

\section{The periodicity-based method}\label{sec:harm}

\subsection{Tunings}

We now present the periodicity-based method and its rationales in detail,
employing relative and logarithmic periodicity. We start with the notion of tuning,
which we define with respect to a system of twelve tones per octave, because in
Western music an octave is usually divided into twelve semitones. 

\begin{defn}\label{def:tun}
A strictly increasing function $\mathcal{T}$ mapping integers (representing
semitone numbers) to positive real numbers (representing frequencies) is called
a {\it tuning}. A tuning is called a \emph{twelve-tone tuning} if it satisfies
$\mathcal{T} (k + 12) = 2 \cdot \mathcal{T}(k)$ for every integer $k$. In this
article we only deal with twelve-tone tunings, so the term \emph{tuning} will
always mean \emph{twelve-tone tuning} in the rest of this article.
\end{defn}

\begin{exmp}\label{exa:12tet}
The frequencies for the $k$th semitone in \emph{equal temperament} with twelve
tones per octave (12-TET) can be computed as $f_k = 2^{k/12} \cdot f$,
where $f$ is the frequency of the ground tone. The respective frequency ratios
$f_k/f$ are shown in the third column of \tabref{tab:tun}. The corresponding tuning function is
$\T(k) = 2^{k/12} \cdot f$. The frequency values grow exponentially and not
linearly, following the \emph{Weber-Fechner law} in psychophysics, which says
that, if the physical magnitude of stimuli grows exponentially, then the
perceived intensity grows only linearly.
\end{exmp}

\begin{table*}
\caption{Table of frequency ratios for different tunings. Here, $k$ denotes the
number of the semitone corresponding to the given interval. In parentheses, the
relative deviation of the respective frequency ratio from equal temperament is
shown. The maximal deviations $d$ for the rational tunings listed here are
1.0\%~(\#1) and 1.1\%~(\#2).}\label{tab:tun}
\footnotesize
\begin{center}\begin{tabular}{lc@{\hspace*{6pt}}c@{}*{4}{r@{/}l@{ }r}} \hline
  Interval & $k$ & Equal temperament & \multicolumn{3}{c}{Pythagorean} & \multicolumn{3}{c}{Kirnberger~III} & \multicolumn{3}{c}{Rational tuning \#1} & \multicolumn{3}{c}{Rational tuning \#2}\\ \hline
  Unison & 0 & 1.000 & 1&1 & (0.00\%) & 1&1 & (0.00\%) & 1&1 & (0.00\%)& 1&1 & (0.00\%)\\
  Minor second & 1 & 1.059 & 256&243 & ($-$0.56\%) & 25&24 & ($-$1.68\%) & 16&15 & (0.68\%) & 16&15 & (0.68\%)\\
  Major second & 2 & 1.122 & 9&8 & (0.23\%) & 9&8 & (0.23\%) & 9&8 & (0.23\%) & 9&8 & (0.23\%)\\
  Minor third & 3 & 1.189 & 32&27 & ($-$0.34\%) & 6&5 & (0.91\%) & 6&5 & (0.91\%) & 6&5 & (0.91\%)\\
  Major third & 4 & 1.260 & 81&64 & (0.45\%) & 5&4 & ($-$0.79\%) & 5&4 & ($-$0.79\%) & 5&4 & ($-$0.79\%)\\
  Perfect fourth & 5 & 1.335 &  4&3 & ($-$0.11\%) & 4&3 & ($-$0.11\%) & 4&3 & ($-$0.11\%) & 4&3 & ($-$0.11\%)\\
  Tritone & 6 & 1.414 & 729&512 & (0.68\%) & 45&32 & ($-$0.56\%) & 17&12 & (0.17\%) & 7&5 & ($-$1.01\%)\\
  Perfect fifth & 7 & 1.498 & 3&2 & (0.11\%) & 3&2 & (0.11\%) & 3&2 & (0.11\%) & 3&2 & (0.11\%)\\
  Minor sixth & 8 & 1.587 & 128&81 & ($-$0.45\%) & 25&16 & ($-$1.57\%) & 8&5 & (0.79\%) & 8&5 & (0.79\%)\\
  Major sixth & 9 & 1.682 & 27&16 & (0.34\%) & 5&3 & ($-$0.90\%) & 5&3 & ($-$0.90\%) & 5&3 & ($-$0.90\%)\\
  Minor seventh & 10 & 1.782 & 16&9 & ($-$0.23\%) & 16&9 & ($-$0.23\%) & 16&9 & ($-$0.23\%) & 9&5 & (1.02\%)\\
  Major seventh & 11 & 1.888 & 243&128 & (0.57\%) & 15&8 & ($-$0.68\%) & 15&8 & ($-$0.68\%) & 15&8 & ($-$0.68\%)\\
  Octave & 12 & 2.000 & 2&1 & (0.00\%) & 2&1 & (0.00\%) & 2&1 & (0.00\%) & 2&1 & (0.00\%)\\ \hline
\end{tabular}\end{center}
\end{table*}

In equal temperament and approximations thereof,
all keys sound more or less equal. This is essential for
playing in different keys on one instrument and for modulation, i.e. changing
from one key to another within one piece of music. Since this seems to be the
practice that is currently in use, at least in Western music, we adopt the equal
temperament with twelve tones as our reference system for tunings.

\begin{rem}\label{NewRemarkOnTPrime}
It is sometimes useful to associate a ratio function $\T'$ to a tuning $\T$. The
\emph{ratio function} $\T'$ is a function from the integers to the positive real
numbers defined by $T'(k)=\T(k)/\T(0)$. Since we assume twelve-tone tuning, the
ratio function consequently satisfies $\T'(k+12)=2\cdot\T'(k)$.
\end{rem}

In equal temperament in \exaref{exa:12tet}, the associated ratio function is
$\T'(k)=2^{k/12}$, as displayed in the third column of \tabref{tab:tun}. The
values of the ratio functions for four other tunings are also displayed
in~\tabref{tab:tun}. 

\subsection{Approximating frequency ratios}\label{sec:approx}

One important feature of the periodicity-based method with smoothing is the use of integer
frequency ratios. In contrast to other approaches, however, we restrict
attention only to integer ratios which differ from the (possibly irrational)
frequency ratios of the real physical signal (with possibly only quasi-periodic
time structure) up to about 1\%. The underlying rationale for approximating the
actual frequency ratios in such a way is that human subjects can distinguish
frequency differences of pure tone components only up to a certain resolution.
It is about 0.7\% at medium and high frequencies under optimal conditions; at
low frequencies, the relative just noticeable difference increases and has a
value of 3.6\% at 100\,Hz \citep{ZF99}. Just intervals, i.e. with integer
frequency ratios, rather than tempered, are considered best in tune, but
everything within a band of $\pm$1\% or even more for more dissonant and
ambiguous intervals, e.g. the tritone, are acceptable to listeners \citep{HH84}.
This also holds, in principle, if the stimuli are pure sinusoids
\citep[Experiment~2]{Vos86} and is in line with the results of \citet{Kop03}
that professional musicians are able to adapt to equal temperament but cannot
really discriminate in their performance between two tuning systems such as
equal temperament and just intonation. In summary, although there are also
interindividual differences and differences between sinusoids and complex tones
as stimuli \citep{ZF99}, it seems to be reasonable to estimate the (relative)
just noticeable difference by 1\% for the musically important low frequency
range, especially the tones in (accompanying) chords \citep{ZFS57,Roe08}.
\opt{long}{The amount of mistuning required to distinguish between a periodic
tone and a tone with one mistuned partial is also in the range of 1\%
\citep[Figure~6.12]{MPG85,MPG86,Har97}.} 

Obviously, the frequency ratios in equal temperament are irrational numbers
(except for the unison and its octaves) but for computing (relative) periodicity (defined in \defref{def:period})
they must be integer fractions, because otherwise no finite period length can be
determined. Let us thus consider tunings with integer frequency ratios. The
oldest tuning with this property is probably the \emph{Pythagorean tuning},
shown in \tabref{tab:tun}. Here, frequency relationships of all intervals have
the form $2^m \cdot 3^n$ for some (possibly negative) integers $m$ and $n$,
i.e. they are based on fifths, strictly speaking, a stack of perfect fifths
(frequency ratio $3/2$) applying octave equivalence. However, although large
numbers appear in the numerators and denominators of the fractions in Pythagorean
tuning, the relative deviations with respect to equal temperament (shown in parentheses
in \tabref{tab:tun}) may be relatively high.\opt{long}{ For example, the tritone
(semitone $6$, frequency ratio 729/512) is still slightly mistuned (deviation
0.68\%). On the other hand, the major third (semitone $4$, frequency ratio
$81/64$, deviation 0.45\%) is tuned more accurately than usually assumed in just
intonation (frequency ratio $5/4$, deviation -0.79\%).} Hence, the
Pythagorean tuning is not in line with the results of psychophysics mentioned
above, that the just noticeable difference of pitch perception is about 1\%.

Let us therefore look for tunings, where the relative deviation $d$ with respect to
equal temperament is approximately~1\% for each tone. In the literature, historical and modern
tunings are listed, e.g. \emph{Kirnberger~III} (cf.~\tabref{tab:tun}). However,
they are also only partially useful in this context, because they do not take
into account the fact on just noticeable differences explicitly. \opt{long}{In
principle, this also holds for the adaptive tunings introduced by \citet{Set05},
where simple integer ratios are used and scales are allowed to vary. An adaptive
tuning can be viewed as a generalised dynamic just intonation, which fits well
with musical practice, because the frequencies for one and the same pitch
category may vary significantly during the performance of a piece of music,
dependent on the musical harmonic context.

}We therefore introduce the \emph{rational tuning} now, which takes the
fractions with smallest possible denominators, such that the relative deviation
with respect to equal temperament is just below a given percentage~$d$.
Frequency ratios $f_i/f$ can be approximated by fractions algorithmically by
employing the so-called Stern-Brocot tree \citep{GKP94,For07}. It induces an
effective procedure for approximating numbers $x$ by fractions $y=a/b$ with some
precision, i.e. maximal relative deviation $d=|y/x-1|$. The main idea is to
perform a binary search between two bounds: $a_l/b_l$ (left) and $a_r/b_r$
(right). We start with the two integer numbers that are nearest to $x$, e.g.
$1/1$ and $2/1$, and repeat computing the so-called \emph{mediant} $a/b =
(a_l+a_r)/(b_l+b_r)$, until $x$ is approximated with the desired precision.
\figref{fig:algo} shows an improved, efficient version of this procedure,
following the lines of~\citet{For07}.

\begin{figure}
\centering
\begin{minipage}{63mm}
\begin{algorithmic}
\STATE $x_{\min} = (1-d)x$;
\STATE $x_{\max} = (1+d)x$;
\STATE $a_l/b_l = \fix{x}/1$;
\STATE $a_r/b_r = (\fix{x}+1)/1$;
\STATE $a/b = [x]/1$;
\WHILE {($a/b < x_{\min}$ \OR $x_{\max} < a/b$)}
\STATE $x_0 = 2 \cdot x-a/b$;
\IF {$(x < a/b)$}
\STATE $a_r/b_r = a/b$;
\STATE $k = \fix{(x_0\,b_l-a_l)/(a_r-x_0\,b_r)}$;
\STATE $a_l/b_l = (a_l+k\,a_r)/(b_l+k\,b_r)$;
\ELSE
\STATE $a_l/b_l = a/b$;
\STATE $k = \fix{(a_r-x_0\,b_r)/(x_0\,b_l-a_l)}$;
\STATE $a_r/b_r = (a_r+k\,a_l)/(b_r+k\,b_l)$;
\ENDIF
\STATE $a/b = (a_l+a_r)/(b_l+b_r)$;
\ENDWHILE
\end{algorithmic}
\end{minipage}
\caption{Approximating real numbers $x$ by fractions $y = a/b$ with maximal
relative deviation $d\opt{long}{= |y/x-1|}$. We use the floor function $\fix{x}$
here, which yields the largest integer less than or equal to $x$, and the
rounding function $[x]$, which rounds $x$ to the nearest integer.\opt{long}{ The
procedure enumerates efficiently, in contrast to continued fraction expansion,
every integer fraction $y = a/b$ with the property that there is no other
fraction with smaller numerator and denominator and smaller deviation $d$ from
$y$ with respect to $x$ \citep[cf.][]{For07}.}}\label{fig:algo}
\end{figure}

\tabref{tab:tun} shows two instances of the rational tuning, namely for $d=1.0$\%
(\#1) \cite[cf.][]{ICMPC.Sto09,Sto10d} and for $d=1.1$\% (\#2). They can be
considered as special just-intonation systems, differing only for the tritone
and the minor seventh. Although relative periodicity values may vary
significantly, if different frequency ratios are used, taking e.g. 16/9 instead
of 9/5 as the frequency ratio for the minor seventh, the evaluation of the
periodicity-based method with smoothing (see~\secref{sec:eval}) shows that the results remain
relatively stable and correlate well with empirical ratings, provided that we
employ the rational tunings \#1 and \#2, which implement that the just
noticeable difference of pitch perception is about 1\%. If we adopt the
frequency ratios e.g. from Pythagorean tuning, which is not psychophysically
motivated in this way, the correlation with empirical ratings of harmonies
decreases.\opt{long}{ \figref{fig:farey} shows possible integer frequency ratios
and their deviation from the nearest semitone in equal temperament.

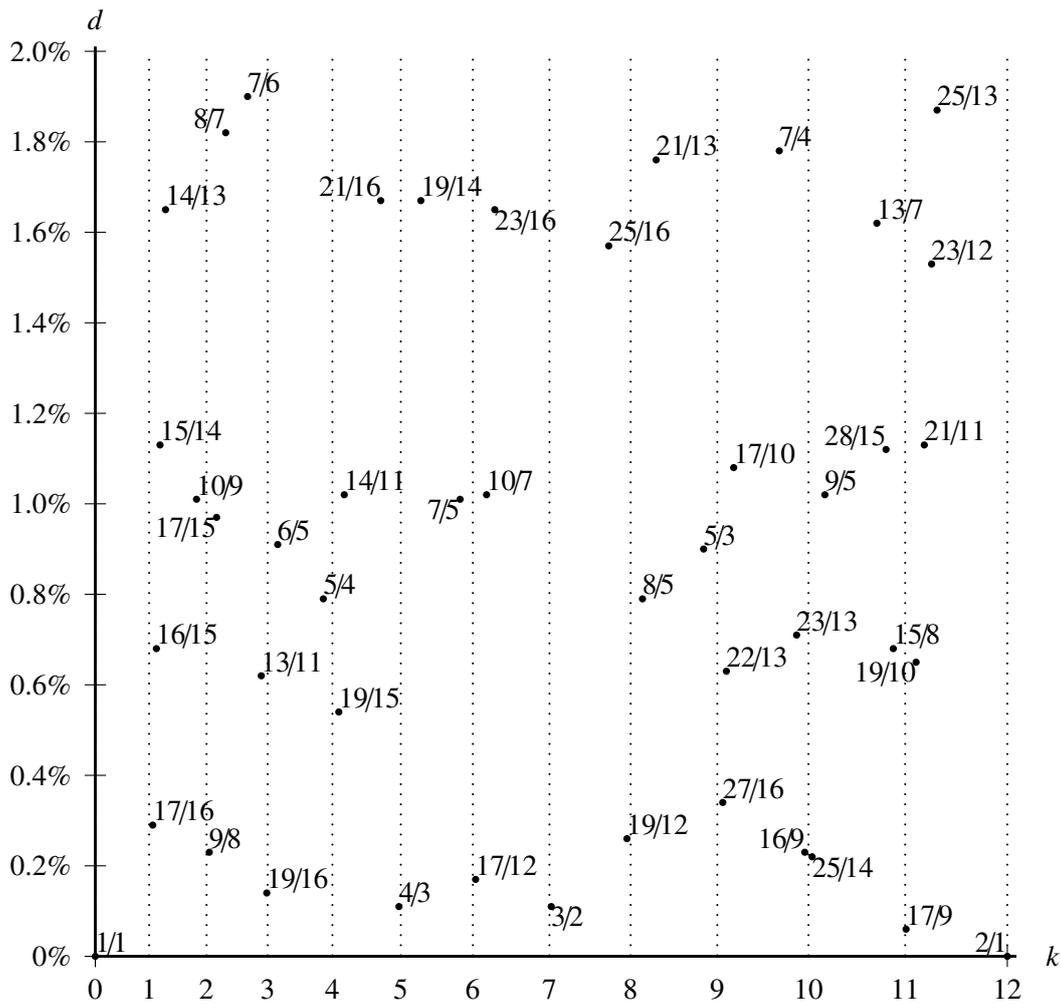
\begin{figure}
\begin{center}
\opt{no.tricks}{\includegraphics[page=2]{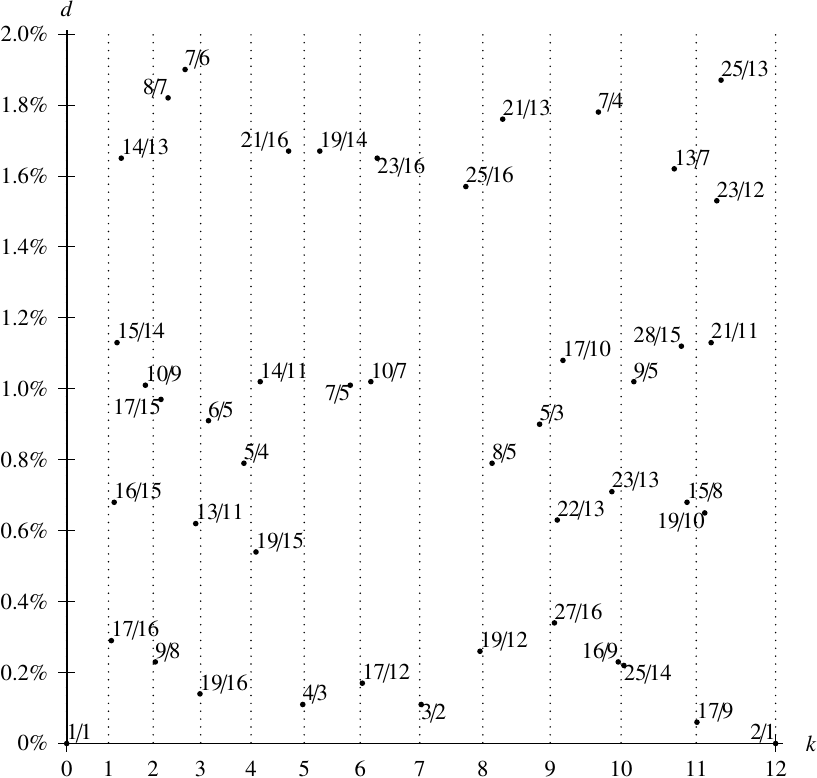}}
\opt{ps.tricks}{
  \psset{xunit=12,yunit=6}
  \renewcommand{\psvlabel}[1]{#1\%}
  \newcommand{\tune}[8][bl]{\qdisk(#2,#8){1.35pt}\rput[#1]{0}(#2,#8){$#3\!/\!#4$}}
  \newcommand{\tone}[2]{\psline[linestyle=dotted]{-}(#2,0)(#2,2)\rput[t]{0}(#2,-0.05){#1}}
\begin{pspicture}(0.9,-0.1)(2.1,2.5)
  \psaxes[labels=y,Dy=0.2,Ox=1]{-}(1,0)(1,0)(2.01,2.01)
  \rput{0}(1,2.07){$d$}
  \rput{0}(2.05,0){$k$}
  \tone{0}{1.000}
  \tone{1}{1.059}
  \tone{2}{1.122}
  \tone{3}{1.189}
  \tone{4}{1.260}
  \tone{5}{1.335}
  \tone{6}{1.414}
  \tone{7}{1.498}
  \tone{8}{1.587}
  \tone{9}{1.682}
  \tone{10}{1.782}
  \tone{11}{1.888}
  \tone{12}{2.000}
  \tune{1.000}{1}{1}{1.000}{0}{+}{0.00}
  \tune[br]{2.000}{2}{1}{2.000}{12}{+}{0.00}
  \tune[tl]{1.500}{3}{2}{1.498}{7}{+}{0.11}
  \tune{1.333}{4}{3}{1.335}{5}{-}{0.11}
  \tune{1.667}{5}{3}{1.682}{9}{-}{0.90}
  \tune{1.250}{5}{4}{1.260}{4}{-}{0.79}
  \tune{1.750}{7}{4}{1.782}{10}{-}{1.78}
  \tune{1.200}{6}{5}{1.189}{3}{+}{0.91}
  \tune[tr]{1.400}{7}{5}{1.414}{6}{-}{1.01}
  \tune{1.600}{8}{5}{1.587}{8}{+}{0.79}
  \tune{1.800}{9}{5}{1.782}{10}{+}{1.02}
  \tune{1.167}{7}{6}{1.189}{3}{-}{1.90}
  \tune[br]{1.143}{8}{7}{1.122}{2}{+}{1.82}
  \tune[bl]{1.429}{10}{7}{1.414}{6}{+}{1.02}
  \tune{1.857}{13}{7}{1.888}{11}{-}{1.62}
  \tune{1.125}{9}{8}{1.122}{2}{+}{0.23}
  \tune{1.875}{15}{8}{1.888}{11}{-}{0.68}
  \tune{1.111}{10}{9}{1.122}{2}{-}{1.01}
  \tune[br]{1.778}{16}{9}{1.782}{10}{-}{0.23}
  \tune{1.889}{17}{9}{1.888}{11}{+}{0.06}
  \tune{1.700}{17}{10}{1.682}{9}{+}{1.08}
  \tune[tr]{1.900}{19}{10}{1.888}{11}{+}{0.65}
  \tune{1.182}{13}{11}{1.189}{3}{-}{0.62}
  \tune{1.273}{14}{11}{1.260}{4}{+}{1.02}
  \tune{1.909}{21}{11}{1.888}{11}{+}{1.13}
  \tune{1.417}{17}{12}{1.414}{6}{+}{0.17}
  \tune{1.583}{19}{12}{1.587}{8}{-}{0.26}
  \tune{1.917}{23}{12}{1.888}{11}{+}{1.53}
  \tune{1.077}{14}{13}{1.059}{1}{+}{1.65}
  \tune{1.615}{21}{13}{1.587}{8}{+}{1.76}
  \tune{1.692}{22}{13}{1.682}{9}{+}{0.63}
  \tune{1.769}{23}{13}{1.782}{10}{-}{0.71}
  \tune{1.923}{25}{13}{1.888}{11}{+}{1.87}
  \tune{1.071}{15}{14}{1.059}{1}{+}{1.13}
  \tune{1.357}{19}{14}{1.335}{5}{+}{1.67}
  \tune[tl]{1.786}{25}{14}{1.782}{10}{+}{0.22}
  \tune{1.067}{16}{15}{1.059}{1}{+}{0.68}
  \tune[tr]{1.133}{17}{15}{1.122}{2}{+}{0.97}
  \tune{1.267}{19}{15}{1.260}{4}{+}{0.54}
  \tune[br]{1.867}{28}{15}{1.888}{11}{-}{1.12}
  \tune{1.063}{17}{16}{1.059}{1}{+}{0.29}
  \tune{1.188}{19}{16}{1.189}{3}{-}{0.14}
  \tune[br]{1.313}{21}{16}{1.335}{5}{-}{1.67}
  \tune[tl]{1.438}{23}{16}{1.414}{6}{+}{1.65}
  \tune{1.563}{25}{16}{1.587}{8}{-}{1.57}
  \tune{1.688}{27}{16}{1.682}{9}{+}{0.34}
\end{pspicture}}
\end{center}
\caption{Integer frequency ratios $a/b$ and their deviation $d$ from the $k$th semitone
in equal temperament for $b \le 16$ and $d \le 2$\%.}\label{fig:farey}
\end{figure}

The approximation procedure for frequency ratios can also be used to
generate (equal temperament) tone scales by an interval, e.g. the perfect fifth.
In this case, we look for a tuning in equal temperament with $n$ tones per octave,
such that the perfect fifth in just intonation (frequency ratio $3/2$) is
approximated as good as possible. Therefore, we develop a fraction $y = a/b$ with
$2^{a/b} \approx 3/2$, where $a$ is the number of the semitone representing the
fifth. Consequently we have to approximate $x = \log_2(3/2) \approx 0.585$ by $y$. The
corresponding sequence of mediants (between $0/1$ and $1/1$) is
$1/2,\;2/3,\;3/5,\;4/7,\;7/12,\;17/29,\;24/41,\;31/53,\;\ldots$ It shows $a/b =
7/12$ among others. Thus, semitone $a=7$ gives the perfect fifth in the tone
scale with $b=12$ tones per octave with high precision. Interestingly, only from
this mediant on, the deviation of $a/b$ with respect to $x$ (which is
logarithmic with respect to the relative frequency) is below 1\%, namely $-0.27$\%.
Hence we obtain 12-TET (see~\exaref{exa:12tet}) by applying once again that the
just noticeable difference between pitches for humans is about 1\%.}

Note that, although we compute the frequency ratios for rational tunings by the
approximation procedure in \figref{fig:algo}, this does not mean that the human
auditory system somehow performs such a computation. We simply assume here,
consistent with the results from psychophysics and neuroacoustics, that the
resolution of periodicity pitch in the brain is limited. The oscillator neurons
in the brain with intrinsic oscillation\opt{long}{ (cf.~\secref{sec:neuronal})}
provide evidence for this. Incidentally, the
time constant $2T = 0.8$\,ms corresponds to a frequency of $1250$\,Hz, which
roughly coincides with the capability of the auditory system to identify the
repetition rate of a complex tone, namely up to about $f=1500$\,Hz
(cf.~\secref{sec:period}). Furthermore, the lowest frequency audible by humans
is about 20\,Hz. Its ratio with the border frequency $f$ is only slightly more
than 1\%. This percentage, corresponding to the above-mentioned just noticeable
difference, is the only parameter of the presented approach on harmony
perception.

\subsection{Harmonies and measures of harmoniousness}

We now define the notions harmony and measure of harmoniousness formally in a
purely mathematical way and give some examples.

\begin{defn}
Assume a finite non-empty set $H = \harm{f_1,\dots,f_k}$ of $k$ positive real
numbers $f_i > 0$ (where $1 \le i \le k$) that represent frequencies. Then $H$ is
called a \emph{harmony} and elements of $H$ are called \emph{tones}. If $f$ is
the minimum of $H$ then $\overline{H} = \{ f_1/f, \dots, f_k/f \}$ is called the
\emph{set of frequency ratios} of the harmony $H$. The set of all harmonies
$H'$ with the same set of frequency ratios is called a \emph{class of
harmonies} and denoted $\clas{\overline{H}} = \{ H' \mid \overline{H'} =
\overline{H} \}$. Furthermore, if all frequency ratios of a harmony are rational
numbers then we say that the harmony and the respective class of harmonies are
\emph{rational}; otherwise we say that they are \emph{irrational}. A
\emph{measure of harmoniousness} $\H$ is a function mapping harmonies to real
numbers such that any two harmonies with the same set of frequency ratios both
map to the same real number. Consequently, $\H$ induces a real-valued function
on the set of harmony classes, which we also denote by $\H$.
\end{defn}

\begin{exmp}\label{ex:zabka}
Consider the harmonies in \tabref{tab:zabka}(a). The harmonies $A_0$ through
$A_4$ represent A major triads in various tuning systems, pitch standards and
musical spacing. The harmony $E$ belongs to the same harmony class as $A_0$ and
represents an E major triad in just intonation under the standard pitch. We see
that all harmonies but $A_3$ are rational. The harmonies $A_0$, $A_1$,
and $E$ belong to the same rational harmony class $\clas{1/1,\,5/4,\,3/2}$. It
is important to note that all harmonies $A_0$ through $A_3$ are notated in the
same way in the usual musical notation.
\end{exmp}

\begin{table*}
\caption{Example harmonies.}\label{tab:zabka}
\renewcommand{\labelenumi}{(\alph{enumi})}
\vspace*{-3mm}
\begin{enumerate}
  \item Major triads in various tuning systems, pitch standards and musical
	spacing. As tunings we here consider rational tuning (\#1 or \#2), the
	Pythagorean tuning, and twelve-tone equal temperament (12-TET). Rational
	tunings \#1 and \#2 yield identical frequency values in the examples
	selected here, so in the Tuning column we do not specify \#1 or \#2
	after the word Rational tuning. Also note that the four entries Rational
	tuning in the Tuning column could also be replaced by Just intonation.
	\begin{center}\footnotesize\begin{tabular}{ccccccc} \hline
	  & Harmony & Rationality & Harmony class & Tuning & Pitch & Spacing\\ \hline
	  $A_0$ & \harm{440,550,660} & Rational & \clas{1/1,5/4,3/2} & Rational tuning & Standard & A4--C$\sharp$5--E5\\
	  $A_1$ & \harm{432,540,648} & Rational & \clas{1/1,5/4,3/2} & Rational tuning & Classical & A4--C$\sharp$5--E5\\
	  $A_2$ & \harm{440,556.875,660} & Rational & \clas{1/1,81/64,3/2} & Pythagorean & Standard & A4--C$\sharp$5--E5\\
	  $A_3$ & $\harm{440,554.365,659.255}$ & Irrational & \clas{1/1,2^{1/3},2^{7/12}} & 12-TET & Standard & A4--C$\sharp$5--E5\\
	  $A_4$ & \harm{440,660,1100} & Rational & \clas{1/1,3/2,5/2} & Rational tuning & Standard & A4--E5--C$\sharp$6\\
	  $E$ & \harm{660,825,990} & Rational & \clas{1/1,5/4,3/2} & Rational tuning & Standard & E5--G$\sharp$5--B5\\ \hline
	\end{tabular}\end{center}
  \item Relative periodicity and gradus suavitatis. The gradus suavitatis can be
	derived from the prime factorisation of the least common multiple (\lcm)
	of the harmonic-series representation. The relative periodicity of the
	respective harmony is simply its minimum.
	\begin{center}\footnotesize\begin{tabular}{clcclccc} \hline
	  & Identifier & Harmonic series & \lcm & Factorisation & Gradus suavitatis & Relative periodicity\\ \hline
	  $A$ & major triad & \harm{4,5,6} & 60 & $2^2 \cdot 3 \cdot 5$ & 9 & 4\\
	  $M_2$ & major second & \harm{8,9} & 72 & $2^3 \cdot 3^2$ & 8 & 8\\
	  $M_3$ & major third & \harm{4,5} & 20 & $2^2 \cdot 5$ & 7 & 4\\
	  $T$ & tritone & \harm{5,7} & 35 & $5 \cdot 7$ & 11 & 5\\ \hline
	\end{tabular}\end{center}
\end{enumerate}
\end{table*}

In \secref{sec:theo}, we have mentioned several measures of harmoniousness,
e.g. gradus suavitatis and relative periodicity. \tabref{tab:zabka}(b) shows their
values for some harmonies. We will come back to this table and explain it in
more detail later, namely after the (formal) definition of relative periodicity
(\defref{def:period}). 

\subsection{Relative periodicity}\label{sec:def}

\begin{figure}
\centering
\begin{minipage}{95mm}
\renewcommand{\labelenumi}{(\alph{enumi})}
\begin{enumerate}
\opt{no.tricks}{
  \item \raisebox{-8mm}{\includegraphics[page=3]{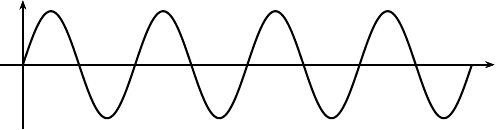}}
  \item \raisebox{-8mm}{\includegraphics[page=4]{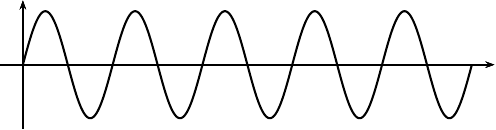}}
  \item \raisebox{-8mm}{\includegraphics[page=5]{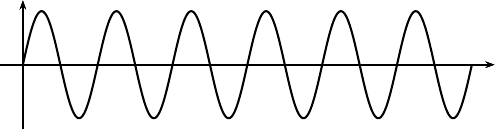}}
  \item \raisebox{-21mm}{\includegraphics[page=6]{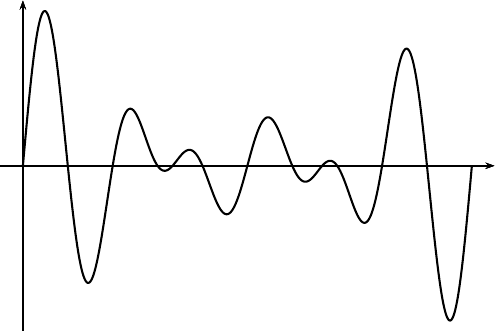}}
  \item \raisebox{-11mm}{\includegraphics[page=7]{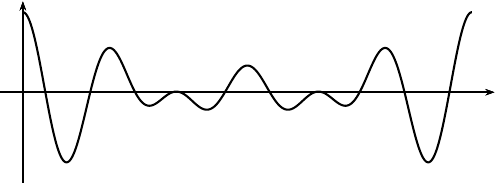}}
  \item \raisebox{-7mm}{\hspace*{9pt}\includegraphics[page=8]{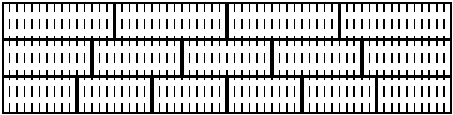}}
}
\opt{ps.tricks}{
\psset{xunit=1.9,yunit=0.9}
\newcommand{\omeg}[1]{\dpi*2^(#1/12)}
  \item \sinusoid{sin(\dpi*1/1*x)}
  \item \sinusoid{sin(\dpi*5/4*x)}
  \item \sinusoid{sin(\dpi*3/2*x)}
  \item \sinusoid[3.1]{sin(\dpi*1/1*x)+sin(\dpi*5/4*x)+sin(\dpi*3/2*x)}
  \item \sinusoid[1.7]{1/2*(cos(\dpi*1/1*x)+cos(\dpi*5/4*x)+cos(\dpi*3/2*x))}
  \item \mbox{}\\[-12pt]\hspace*{9pt}
\begin{pspicture}(0,0)(4,2.1)
  \psframe(0,1.4)(1,2.1)\psframe(1,1.4)(2,2.1)\psframe(2,1.4)(3,2.1)\psframe(3,1.4)(4,2.1)
  \psframe(0,0.7)(0.8,1.4)\psframe(0.8,0.7)(1.6,1.4)\psframe(1.6,0.7)(2.4,1.4)\psframe(2.4,0.7)(3.2,1.4)\psframe(3.2,0.7)(4,1.4)
  \psframe(0,0)(0.667,0.7)\psframe(0.667,0)(1.333,0.7)\psframe(1.333,0)(2,0.7)\psframe(2,0)(2.667,0.7)\psframe(2.667,0)(3.333,0.7)\psframe(3.333,0)(4,0.7)
  \psset{linestyle=dashed,dotsep=4pt}
  \psline(0.067,0)(0.067,2.1)\psline(1.067,0)(1.067,2.1)\psline(2.067,0)(2.067,2.1)\psline(3.067,0)(3.067,2.1)
  \psline(0.133,0)(0.133,2.1)\psline(1.133,0)(1.133,2.1)\psline(2.133,0)(2.133,2.1)\psline(3.133,0)(3.133,2.1)
  \psline(0.2,0)(0.2,2.1)\psline(1.2,0)(1.2,2.1)\psline(2.2,0)(2.2,2.1)\psline(3.2,0)(3.2,2.1)
  \psline(0.267,0)(0.267,2.1)\psline(1.267,0)(1.267,2.1)\psline(2.267,0)(2.267,2.1)\psline(3.267,0)(3.267,2.1)
  \psline(0.333,0)(0.333,2.1)\psline(1.333,0)(1.333,2.1)\psline(2.333,0)(2.333,2.1)\psline(3.333,0)(3.333,2.1)
  \psline(0.4,0)(0.4,2.1)\psline(1.4,0)(1.4,2.1)\psline(2.4,0)(2.4,2.1)\psline(3.4,0)(3.4,2.1)
  \psline(0.467,0)(0.467,2.1)\psline(1.467,0)(1.467,2.1)\psline(2.467,0)(2.467,2.1)\psline(3.467,0)(3.467,2.1)
  \psline(0.533,0)(0.533,2.1)\psline(1.533,0)(1.533,2.1)\psline(2.533,0)(2.533,2.1)\psline(3.533,0)(3.533,2.1)
  \psline(0.6,0)(0.6,2.1)\psline(1.6,0)(1.6,2.1)\psline(2.6,0)(2.6,2.1)\psline(3.6,0)(3.6,2.1)
  \psline(0.667,0)(0.667,2.1)\psline(1.667,0)(1.667,2.1)\psline(2.667,0)(2.667,2.1)\psline(3.667,0)(3.667,2.1)
  \psline(0.733,0)(0.733,2.1)\psline(1.733,0)(1.733,2.1)\psline(2.733,0)(2.733,2.1)\psline(3.733,0)(3.733,2.1)
  \psline(0.8,0)(0.8,2.1)\psline(1.8,0)(1.8,2.1)\psline(2.8,0)(2.8,2.1)\psline(3.8,0)(3.8,2.1)
  \psline(0.867,0)(0.867,2.1)\psline(1.867,0)(1.867,2.1)\psline(2.867,0)(2.867,2.1)\psline(3.867,0)(3.867,2.1)
  \psline(0.933,0)(0.933,2.1)\psline(1.933,0)(1.933,2.1)\psline(2.933,0)(2.933,2.1)\psline(3.933,0)(3.933,2.1)
  \psline(1,0)(1,2.1)\psline(2,0)(2,2.1)\psline(3,0)(3,2.1)\psline(3.8,0)(3.8,2.1)
\end{pspicture}
}
\end{enumerate}
\end{minipage}
\caption{Sinusoids of the major triad in root position (a)--(c), their
superposition (d), and corresponding autocorrelation function $\rho(\tau)$ in (e).
\figref{fig:major}(f) shows the general periodicity structure of the chord,
abstracting from concrete overtone spectra and amplitudes. \opt{long}{The solid
boxes show  the periodic patterns of the three tone components one upon the
other. The dashed lines therein indicate the greatest common period of all tone
components, called \emph{granularity} by \citet{ICMPC.Sto12b}. Its frequency
corresponds to their least common overtone.}}\label{fig:major}
\end{figure}
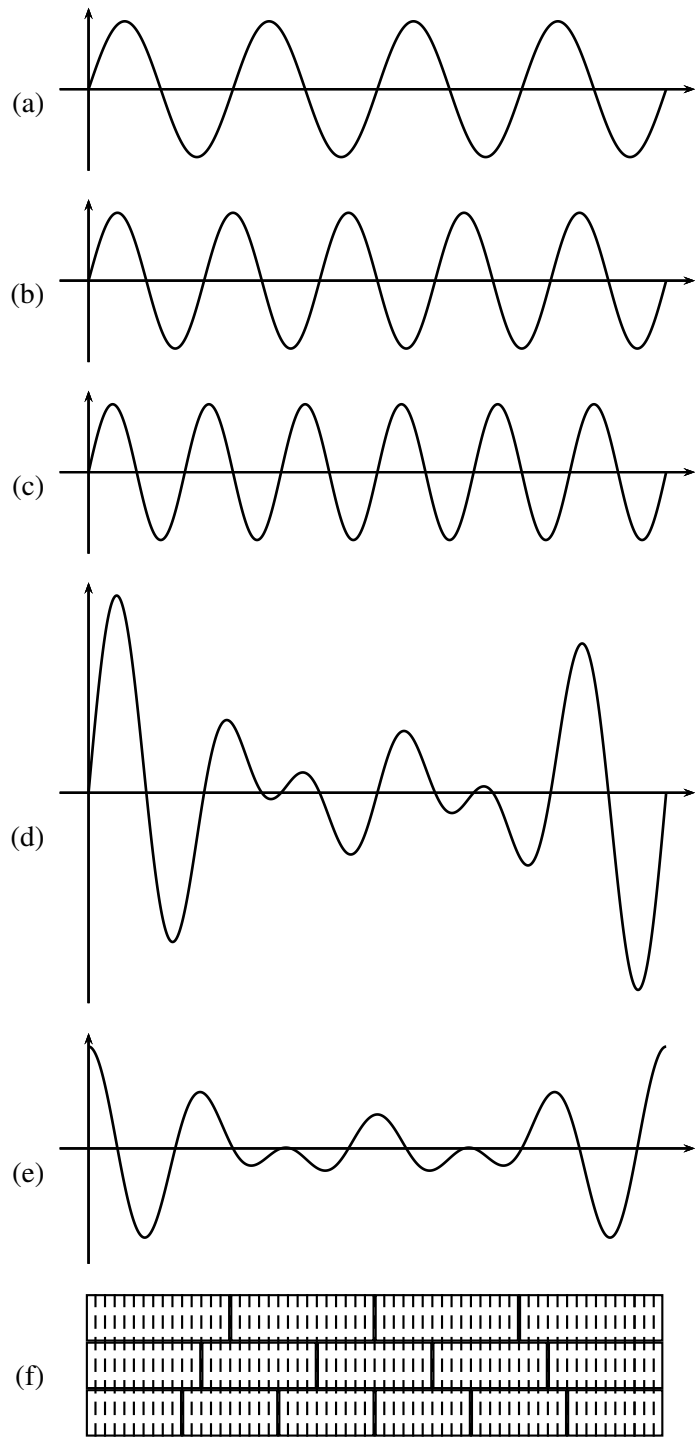

We now introduce the concept of relative periodicity, at first informally with an
example: For this, we further consider the harmony $A_0$ from \tabref{ex:zabka}(a),
i.e. the A~major triad in root position (cf.~\figref{fig:triad}(a)) and in
just intonation. It consists of three tones with absolute frequencies
$f_1=440$\,Hz (lowest tone), $f_2=550$\,Hz (major third), and $f_3=660$\,Hz
(perfect fifth). For the sake of simplicity, we ignore possible overtones and
consider just the three pure tone components here. \figref{fig:major}(a)--(c) shows
the sinusoids for the three pure tone components and \figref{fig:major}(d) their
superposition, i.e. the graph of the function $s(t) = \sin(\omega_1t) +
\sin(\omega_2t) + \sin(\omega_3t)$, where $\omega_i = 2\pi f_i$ is the
respective angular frequency. 

How can the periodicity of the signal $s(t)$ in \figref{fig:major}(d) be
determined? One possibility is to apply \emph{continuous autocorrelation}, i.e.
the cross-correlation of a signal with itself. For the superposition of periodic
functions $s(t)$ over the reals, it is defined as the continuous cross-correlation
integral of $s(t)$ with itself, at lag $\tau$, as follows: \[ \rho(\tau) = \lim_{T
\to \infty} \frac{1}{2\,T} \int_{-T}^{+T} s(t) \cdot s(t-\tau) \,\mathrm{d}t = \frac{1}{2} \sum_{i=1}^3
\cos(\omega_i \tau) \] The autocorrelation function reaches its peak at the
origin. Other maxima indicate possible period lengths of the original signal.
Furthermore, possibly existing phase shifts are nullified, because we always
obtain a sum of pure cosines with the same frequencies as in the original signal
as above. The respective graph of $\rho(\tau)$ for the major triad example is
shown in \figref{fig:major}(e). As one can see, it has a peak after four times the
period length of the lowest tone (cf.~\figref{fig:major}(a)). This corresponds to the
periodicity of the envelope frequency, to which respective neurons in the inner
cortex respond.

In general, overtones have to be taken into account in the computation of the
autocorrelation function, because real tones have them. In many approaches,
complex tones are made up from sinusoidal partials in this context. Their
amplitudes vary e.g. as $1/k$, where $k$ is the number of the partial, ranging
from 1 to 10 or similar \citep{HK78,HK79,Set05,Ebe07,Ebe08}, although the number
of partials can be higher in reality, if one looks at the spectra of musical instruments.
In contrast to this procedure, we calculate relative periodicity, abstracting from concrete overtone
spectra, by simply considering the frequency ratios of the involved tones.
\figref{fig:major}(f) illustrates this by showing the periodic patterns of the
three tone components of the major triad as solid boxes one upon the other. As
one can see, the period length of the chord is (only) four times the period
length of the lowest tone for this example. This ratio is the relative
periodicity $h$. It only depends on the corresponding harmony class, which must
be rational, as already stated in~\secref{sec:approx}.

For a harmony class \clas{a_1/b_1, \ldots, a_k/b_k} (where all fractions
$a_i/b_i$ are in lowest terms, i.e. $a_i$ and $b_i$ are coprime), the value
of $h$ can be computed as $\lcm(b_1, \ldots, b_k)$, i.e., it is the least common
multiple (\lcm) of the denominators of the frequency ratios. This can be seen as
follows: Since the relative period length of the lowest tone $T_1 = f_1/f_1$ is
$1$, we have to find the smallest integer number that is an integer multiple of
all relative period lengths $T_i=f_1/f_i = b_i/a_i$ for $1 < i \le k$. Since
after $a_i$ periods of the $i$th tone, we arrive at the integer $b_i$, $h$ can
be computed as the least common multiple of all $b_i$. For the harmony class
\clas{1/1,5/4,3/2} of $A_0$ e.g., we obtain $a_2 \cdot T_2 = 3 \cdot 2/3 = 2 = b_2$
for $T_2 = f_1/f_2 = 2/3$. Together with $b_3 = 4$, ignoring $b_1 = 1$, which
is always irrelevant, we get $h = \lcm(1,2,4) = 4$ as expected. Let us now capture the
notion of relative periodicity formally.

\begin{defn}\label{def:period}
Assume that $H = \harm{f_1,\dots,f_k}$ is a rational harmony, $f$ is the
minimum of $H$, and $f_i/f = a_i/b_i$ for $1 \le i \le k$ and coprime
positive integers $a_i$ and $b_i$. Denote $h = \lcm(b_1,\dots,b_k)$. Then $h$ is
called \emph{relative periodicity} and the set $H^\mathrm{harm} = h\overline{H}
= \harm{ h \cdot a_i/b_i \mid 1 \le i \le k }$ is called the
\emph{harmonic-series representation} of the rational harmony $H$.
\end{defn}

\begin{exmp}\label{ex:period}
The set of frequency ratios for the harmonies $A_0$, $A_1$, and $E$ from
\tabref{ex:zabka}(a) is \harm{1/1,5/4,3/2}. Thus, as already mentioned, their
relative periodicity is $4$ and their harmonic-series representation is
\harm{4,5,6}. The relative periodicity of $A_2$ is $64$ and its harmonic-series
representation is \harm{64,81,96}.
\end{exmp}

Note that in all cases of \exaref{ex:period}, the relative periodicity is
the minimal element of the harmonic-series representation. It can easily be
shown that this is not a coincidence but a consequence of \defref{def:period}.
It always holds that $h = \min(H^\mathrm{harm})$. This equation may be
used as an alternative way of defining the key concept: Relative periodicity
can be defined as the minimum of the harmonic-series representation of a given
rational harmony.

Interestingly, both relative periodicity and gradus suavitatis are functions of the
harmonic-series representation: While relative periodicity corresponds to its
minimum, the gradus suavitatis depends on the least common multiple of the
harmonic-series representation\opt{long}{ (cf.~\secref{sec:ratio})}. Thus, relative periodicity only considers the
lowest element in the harmonic-series representation, whereas Euler's formula
takes into account more complex harmonic relations of all its elements. Some
values of both measures of harmoniousness are listed in \tabref{tab:zabka}(b).
Note that on the one hand relative periodicity may not change for harmonies of
increasing complexity: For instance, relative periodicity is the same for the 
major triad in root position and the major third (which is contained in the major triad) in just intonation (called
$A$ and $M_3$ in \tabref{tab:zabka}(b)), while their gradus suavitatis values
differ. On the other hand, both measures of harmoniousness produce different
orderings: For example, relative periodicity predicts the preference tritone
$\prec$ major second (called $T$ and $M_2$ in \tabref{tab:zabka}(b)) which
corresponds to the empirical ranking (cf.~\tabref{tab:dyad}), while it is
the other way round for the gradus suavitatis: We have $h(T)=5 < h(M_2)= 8$, but
$\Gamma(M_2)=8 < \Gamma(T)=11$.

\subsection{A hypothesis on harmony perception}

The rationale behind employing relative periodicity as a measure of
harmoniousness is given by the recent results from neuroacoustics on periodicity
detection in the brain, which we already reviewed in \secref{sec:neuronal}.
Clearly, different periodicity pitches must be mapped on different places on the
periodicity map. However, a general observation is that harmony perception is mostly independent of transposition, i.e. pitch shifts.
For instance, the perceived consonance/dissonance of an A~major triad should be
more or less the same as that of one in~B. This seems to be in accordance with
the logarithmic organisation of the neuronal periodicity map. There is no
problem with the fact that periodicity detection in the brain is not available
for the high frequency range (cf.~\secref{sec:period}),
because the fundamental tones (not necessarily the overtones) of real musical
harmonies usually are below this threshold, and this suffices to determine
periodicity.

In line with \citet{MM+13}, one may conclude that harmony perception requires in addition some cognitive process or property
filter \citep{Lic62}, conveying information concerning stimulus periodicities in
(short-term) memory. The latter is also required to explain harmony perception
when tones occur consecutively. Because of this, we do not consider
absolute period length here, but relative periodicity. As we shall see (in
\secref{sec:eval}), this measure shows very high correlation with empirical
ratings of harmonies, especially higher than that of roughness \citep{HK78,HK79}
which only regards the auditory processing in the ear and not the brain. This is
in line with the result of \citet{CMP12} that the quality of roughness (induced
by beating) constitutes an aesthetic dimension that is distinct from that of
dissonance.

It is a feature of the periodicity-based method with smoothing that the actual amplitudes
of the tone components in the given harmony are ignored. Harmonic overtone
spectra are irrelevant for determining relative periodicities, because the
period length of a waveform of a complex tone with harmonic overtones is
identical with that of its fundamental tone. We always obtain $h=1$, since the
frequencies of harmonic overtones are integer multiples of the fundamental
frequency. All frequency ratios \clas{1/1,2/1,3/1,\ldots} have $1$ as
denominator in this case.
Therefore, relative periodicity $h$ is independent of concrete amplitudes and
also phase shifts of the pure tone components, i.e. tones with a plain sinusoidal
waveform. Incidentally, information on the waveform of the original signal is
lost in the auditory processing, because neuronal activity usually has the form
of spike trains \citep{LS_88,Car99,TC+01}. Information on phase is also lost
because of the phase-locking mechanism in the brain \citep{LS_88,MH91,LS+09}.
However, the information on periodicity remains, which is consistent with the
periodicity-based method proposed here.

\opt{long}{Also from a more practical point of view, it seems plausible that
harmony perception depends more on periodicity than on loudness and timbre of
the sound. It should not matter much whether a chord is played e.g. on guitar,
piano, or pipe organ. Of course, this argument only holds for tones with
harmonic overtone spectra. If we have inharmonic overtones in a complex tone
such as in Indian, Thai, or Indonesian gong orchestra, i.e. Gamelan
music, or stretched or compressed timbres as considered by \citet{Set05}, then
it holds $h>1$ for the relative periodicity value of a single tone, i.e., we
have an inherently increased \emph{harmonic complexity} \citep{Par89}. Average
listeners seem to prefer low or middle harmonic complexity, e.g. baroque or
classical style on the one hand and impressionism and jazz on the other hand,
which of course have to be analysed as complex cultural phenomena. In this
context, \citet[pp.\,57-58]{Par89} speaks of \emph{optimal dissonance} that
gradually increased during the history in Western music.}

We therefore set up the following \emph{hypothesis on harmony perception}:
The perceived consonance of a harmony decreases as the relative periodicity $h$ increases.
For the major triad in root position, we have $h=4$ (cf.~\exaref{ex:period}), which is
quite low. Thus, the predicted consonance is high. This correlates very well with
empirical results. Relative periodicity gives us a powerful approach to the
analysis of musical harmony perception. We evaluate this hypothesis later at
length (\secref{sec:eval}). But beforehand, we will state some refinement of
relative periodicity.

\subsection{Logarithmic periodicity and smoothing}

Throughout the rest of this article, we consider relative periodicity $h$ as
measure of harmoniousness and also its base-2 logarithm $\log_2(h)$, called
\emph{logarithmic periodicity} henceforth. The rationale for taking the
logarithm of relative periodicity is given again by the recent results from
neuroacoustics on periodicity detection in the brain, namely the logarithmic
organisation of the neuronal periodicity map in the brain
(cf.~\secref{sec:neuronal}). Since one octave corresponds to a frequency ratio
of $2$, we adopt the base-2 logarithm. Obviously, as $\log_2$ is an increasing
function, logarithmic periodicity and relative periodicity lead to identical
harmony rankings. This changes, however, if we apply the concept of smoothing
(see~\defref{def:smooth}). For this, harmonies must be given by sets of
semitones, as follows.

\begin{defn}
A \emph{set of semitones} $S = \semi{s_1,\dots,s_n}$ is a set of (possibly
negative) integer numbers, usually containing $0$. Such a set $S$ may be \emph{shifted} by
$i$ semitones, defined by $S_i = \semi{s_1-i, \dots, s_n-i}$. If $S$ is a set of
semitones and $\T'$ a ratio function, we write $\T'(S)$ for $\{\T'(s_1),
\dots, \T'(s_n)\}$.
\end{defn}

\begin{defn}\label{def:smooth}
Let $\T'$ be a ratio function, $S = \semi{s_1,\dots,s_n}$ a set of $n$
semitones, and $\H$ a measure of harmoniousness. Then, the value of $\H$ may be
\emph{smoothed}, obtaining the (smoothed) measure of harmoniousness
$\overline{\H}$ by averaging over the shifted semitone sets of $S$, as follows:
\[ \overline{\H}(S) = \frac{1}{n} \underset{i \in S}{\sum} \H(\T'(S_i)) \]
\end{defn}

Most comparisons with empirical rankings yield comparably favourable results, if
(raw) relative periodicity is employed, which can be computed without reference
to any tuning system. However, in some cases there is more than one integer
fraction approximating a given frequency ratio with the required precision of about
1\%. Therefore, for harmonies given as semitone sets, smoothing averages over
several related periodicity values. Only the tones of the given harmony are used
as reference tones in the procedure according to \defref{def:smooth}, because
otherwise the frequency ratios often deviate more than 1\%. For instance, for
the perfect fifth with the set of semitones $S = \semi{0,7}$ we do not consider e.g.
the shifted set $S_{10} = \semi{-10,-3}$ which corresponds to the frequency
ratios \harm{9/16,5/6} in rational tuning (\#1 and \#2). The quotient of both frequency ratios is
$40/27$, which not only deviates from the desired frequency ratio of the
perfect fifth by more than 1\% but also has an unnecessarily large denominator
(that corresponds to its relative periodicity). 

\begin{exmp}
Let us consider the first inversion of the diminished triad consisting of the
tones A4, C5, and F$\sharp$5 (see \figref{fig:triad}(c)). The corresponding
set of semitones is $S = S_0 = \semi{0,3,9}$, where the lowest tone has the
number $0$ and is associated with the frequency ratio $1/1$ (unison). We use the
corresponding frequency ratios according to rational tuning \#2, i.e. $\T'(S) =
\harm{1/1,6/5,5/3}$. Hence, its relative periodicity is $h_0 = \lcm(1,5,3) =
15$. In order to smooth and hence stabilise the calculated periodicity value, we
have to consider also the shifted sets of semitones $S_3 = \semi{-3,0,6}$ and
$S_9 = \semi{-9,-6,0}$. As a consequence of \defref{def:tun}, for semitones
associated with a negative number $-n$, we take the frequency ratio of semitone
$12-n$ and halve it, i.e., we do not assume octave equivalence here.
Therefore, we get the frequency ratios $\T'(S_3) = \{5/6,1/1,7/5\}$ and
$\T'(S_9) = \{3/5,7/10,1/1\}$ with relative periodicity values $h_3 = 5/6 \cdot
\lcm(6,1,5) = 25$ and $h_9 = 3/5 \cdot \lcm(5,10,1) = 6$, respectively. Their
arithmetic average and hence the smoothed relative periodicity is $\overline{h}
= (15+25+6)/3 \approx 15.3$. The smoothed logarithmic periodicity in this case
is $\overline{\log_2(h)} = (\log_2(15) + \log_2(25) + \log_2(6))/3 \approx 3.7$.
\end{exmp}

Note that, as a side effect of the procedure, for logarithmic periodicity we
implicitly obtain the geometric average over all $h$ values instead of the
arithmetic average. Because of the Weber-Fechner law and the logarithmic mapping
of frequencies on both the basilar membrane in the inner ear and the periodicity
map in the brain, this seems to be entirely appropriate.

\begin{exmp}\label{exa:spread}
As a further example, we consider once again the major triad, but this time spread
over more than one octave, consisting of the tones C3 (lowest tone), E4
(major tenth), and G4 (perfect twelfth) with corresponding numbers of
semitones \semi{0,16,19}.
In contrast to other approaches, we do not project the tones into one octave,
but apply a factor $2$ for each octave. Thus by \remref{NewRemarkOnTPrime}, the
frequency ratio for the major tenth with corresponding semitone number $16$ is
$\T'(16)=2 \cdot \T'(4) =2 \cdot 5/4 = 5/2$ in lowest terms.
The frequency ratios of
the whole chord are \harm{1/1,5/2,3/1}. Hence, $h=\lcm(1,2,1) = 2$ and
$\log_2(h) = 1$. Smoothing does not change the results here.
\end{exmp}

\opt{long}{Perceived consonance/dissonance certainly depends on overtones to a
certain extent. Since concrete amplitudes are neglected in the periodicity-based
method, there are ambiguous cases where harmonic overtones cannot be
distinguished from extra tones coinciding with these overtones. This holds for
\exaref{exa:spread}, because it contains the tones C3 and G4, which are more
than an octave apart, namely $n=19$ semitones (perfect twelfth), with frequency
ratio $3/2 \cdot 2= 3/1$. It cannot be distinguished from a complex tone with C3
as fundamental tone and harmonic overtones, because its third partial
corresponds to G4. However, the waveforms normally differ here: In the former
case the amplitudes of C3 and G4 may be almost equal, whereas in the latter case
the amplitude $a$ of the overtone G4 may be significantly lower than that of C3.
The superposition of both sinusoids can be stated roughly as $\sin(\omega t) +
a\sin(3\omega t)$, where $\omega = 2\pi f$ with $f \approx 131$\,Hz (the
frequency of C3), and $t$ is the time. For $-1/3 \le a \le 1/9$, the higher tone
component G4 does not induce any additional local extrema in the waveform, which
would correspond to additional spikes in the neuronal activity. It seems that
people tend to underestimate the number of pitches in such chords
\citep[p.~5]{DC87,MM+13}. Nonetheless, even in such ambiguous cases, the
periodicity-based method yields meaningful results and high correlations to
empirical ratings of harmony perception for realistic chords
(see~\secref{sec:triad}).}

\opt{long}{\begin{exmp}
Let us now consider the minor triad in root position. From the respective frequency
ratios in rational tuning (\#1 or \#2) $\harm{1/1, 6/5, 3/2}$, we easily
can read off the relative periodicity $h = \lcm(1, 5, 2) = 10$. Smoothing does
not change the result here. Here it is important to note that pitch and
periodicity are orthogonal dimensions of harmony perception on the neuronal
tonotopic map (cf.~\secref{sec:neuronal}). If we ignore this and identify tone
pitch and periodicity pitch, we obtain e.g. for the A minor triad in root
position (cf.~\figref{fig:triad}(b)) the tone F2 as missing fundamental (with
$1/h = 1/10$ of the lowest frequency in the harmony) that does not at all belong
to the chord.
\end{exmp}}

\begin{exmp}
Let us finally calculate the periodicity of the complete
chromatic scale, i.e. the twelve tones constituting Western music. For this, we
compute the least common multiple of the denominators of all frequency
ratios according to rational tuning \#2 within one octave (cf.~\tabref{tab:tun}). We
obtain $h = \lcm(1,15,8,5,4,3,5,2,5,3,5,8) = 120$. Smoothing yields
$\overline{h} \approx 168.2$ and $\overline{\log_2(h)} \approx
7.4$. Thus interestingly, the (smoothed) logarithmic periodicity for the chromatic scale is
just within the biological bound of $8$ octaves, which can be represented in the
neuronal periodicity map (cf.~\secref{sec:neuronal}).
\end{exmp}

\section{Results and evaluation}\label{sec:eval}

Let us now apply the periodicity-based method and other theories on harmony
perception to common musical harmonies and correlate the obtained results with
empirical results \citep{Mal18,Rob86,SHP03,JKL12,TT13}. The corresponding
experiments are mostly conducted by (cognitive) psychologists, where the
harmonies of interest are presented singly or in context. The listeners are
required to judge the consonance/dissonance, using an ordinal scale. All
empirical and theoretical consonance values are either taken directly from the
cited references or calculated by computer programs according to the respective
model on harmony perception. In particular, the smoothed periodicity values
$\overline{h}$ and $\overline{\log_2(h)}$ are computed by a program, implemented
by the author of this article, written in the declarative programming language
ECLiPSe~Prolog \citep{AW07,CM10}. A table listing the computed harmoniousness values for all 
$2048$ possible harmonies within one octave consisting of up to $12$ semitones
among other material is available at \url{http://artint.hs-harz.de/fstolzenburg/harmony/},
including a related technical report version of this article \citep{Sto13}.

In our analyses, we correlate the empirical and the theoretical ratings of
harmonies. Since in most cases only data on the ranking of harmonies is
available, we mainly correlate rankings. Nevertheless, correlating concrete
numerical values yields additional interesting insights (see
\secref{sec:triad}). For the sake of simplicity and consistency, we always
compute Pearson's correlation coefficient $r$, which coincides with Spearman's
rank correlation coefficient on rankings, provided that there are not too many
bindings, i.e. duplicate values. For determining the significance of the
results, we apply $r \sqrt{n-2} / \sqrt{1-r^2} \sim t_{n-2}$, i.e. we have
$n-2$ degrees of freedom in Student's $t$ distribution, where $n$ is the number
of corresponding ranks or values. We always perform a one-sided test whether $r
\not> 0$.

\subsection{Dyads}

\tabref{tab:dyad} shows the perceived and computed consonance of dyads
(intervals). The empirical rank is the average ranking according to the summary
given by \citet[Figure~6]{SHP03}, which includes the data by \citet{Mal18}.
\tabref{tab:cor2} provides a more extensive list of approaches on
harmony perception, indicating the correlation of the rankings together with its
significance. As one can see, the correlations of the empirical rating with the
sonance factor and with smoothed relative\opt{long}{ and smoothed logarithmic} periodicity show the
highest correlation ($r=.982$).

\begin{table*}
\caption{Consonance rankings of dyads. The respective sets of semitones are
given in braces, raw values of the
respective measures in parentheses. The empirical rank is the average rank
according to the summary given by \citet[Figure~6]{SHP03}. The roughness values
are taken from \citet[Appendix]{HK78}. For computing the sonance factor
\citep{Hof04,Hof08}, the \emph{Harmony Analyzer 3.2} applet software has been
used, available at \protect\url{http://www.chameleongroup.org.uk/software/piano.html}.
For these models, always C4 (middle C) is taken as the lowest tone. For smoothed relative periodicity and percentage
similarity \citep{GP09}, the frequency ratios from rational tuning \#2 are used.}\label{tab:dyad}
\footnotesize
\centering
\begin{tabular}{l@{ }rc*{4}{c@{\,}c}} \hline
  & & Empirical & & & & & & & \multicolumn{2}{c}{Smoothed relative} \\
  \multicolumn{2}{l}{Interval} & rank & \multicolumn{2}{c}{Roughness} & \multicolumn{2}{c}{Sonance factor} & \multicolumn{2}{c}{Similarity} & \multicolumn{2}{c}{periodicity}\\ \hline
  Unison	& \semi{0,0}	& 1	& 2	& (0.0019)	& 1-2	& (1.000)	& 1-2	& (100.00\%)	& 1-2	& (1.0)\\
  Octave	& \semi{0,12}	& 2	& 1	& (0.0014)	& 1-2	& (1.000)	& 1-2	& (100.00\%)	& 1-2	& (1.0)\\
  Perfect fifth	& \semi{0,7}	& 3	& 3	& (0.0221)	& 3	& (0.737)	& 3	& (66.67\%)	& 3	& (2.0)\\
  Perfect fourth& \semi{0,5}	& 4	& 4	& (0.0451)	& 4	& (0.701)	& 4	& (50.00\%)	& 4-5	& (3.0)\\
  Major third	& \semi{0,4}	& 5	& 6	& (0.0551)	& 5	& (0.570)	& 6	& (40.00\%)	& 6	& (4.0)\\
  Major sixth	& \semi{0,9}	& 6	& 5	& (0.0477)	& 6	& (0.526)	& 5	& (46.67\%)	& 4-5	& (3.0)\\
  Minor sixth	& \semi{0,8}	& 7	& 7	& (0.0843)	& 7	& (0.520)	& 9	& (30.00\%)	& 7-8	& (5.0)\\
  Minor third	& \semi{0,3}	& 8	& 10	& (0.1109)	& 8	& (0.495)	& 7	& (33.33\%)	& 7-8	& (5.0)\\
  Tritone	& \semi{0,6}	& 9	& 8	& (0.0930)	& 11	& (0.327)	& 8	& (31.43\%)	& 9	& (6.0)\\
  Minor seventh	& \semi{0,10}	& 10	& 9	& (0.0998)	& 9	& (0.449)	& 10	& (28.89\%)	& 10	& (7.0)\\
  Major second	& \semi{0,2}	& 11	& 12	& (0.2690)	& 10	& (0.393)	& 11	& (22.22\%)	& 12	& (8.5)\\
  Major seventh	& \semi{0,11}	& 12	& 11	& (0.2312)	& 12	& (0.242)	& 12	& (18.33\%)	& 11	& (8.0)\\
  Minor second	& \semi{0,1}	& 13	& 13	& (0.4886)	& 13	& (0.183)	& 13	& (12.50\%)	& 13	& (15.0)\\ \hline
  \multicolumn{2}{l}{Correlation~$r$} & & .967 & & .982 & & .977 & & .982 & \\ \hline
\end{tabular}
\end{table*}

\begin{table*}
\caption{Correlations of several consonance rankings with empirical ranking for
dyads. For percentage similarity \citep[Table~1]{GP09}, gradus suavitatis
\citep{Eul39}, consonance value \citep{Bre05}, and the smoothed $\Omega$ measure
\citep{ICMPC.Sto12b}, the frequency ratios from rational tuning \#2 (see
\tabref{tab:tun}) are used.}\label{tab:cor2}
\footnotesize
\begin{center}\begin{tabular}{lcc} \hline
  Approach & Correlation $r$ & Significance $p$\\ \hline
  Sonance factor \citep{Hof04,Hof08} & .982 & .0000\\
  Smoothed relative and smoothed logarithmic periodicity (rational tuning \#2) & .982 & .0000\\
  Consonance raw value \citep[Figure~5]{ICMPC.Fol12} & .978 & .0000\\
  Percentage similarity \citep[Table~1]{GP09} & .977 & .0000\\
  Roughness \citep[Appendix]{HK78} & .967 & .0000\\
  Smoothed gradus suavitatis \citep{Eul39} & .941 & .0000\\
  Consonance value \citep{Bre05} & .940 & .0000\\
  Pure tonalness \citep[p.\,140]{Par89} & .938 & .0000\\
  Smoothed relative and smoothed logarithmic periodicity (rational tuning \#1) & .936 & .0000\\
  Dissonance curve \citep[Figure~6.1]{Set05} & .905 & .0000\\
  Smoothed $\Omega$ measure \citep{ICMPC.Sto12b} & .886 & .0000\\
  Generalised coincidence function \cite[Figure~3B]{Ebe08} & .841 & .0002\\
  Smoothed relative periodicity (Pythagorean tuning) & .817 & .0003\\
  Smoothed relative periodicity (Kirnberger~III) & .796 & .0006\\
  Complex tonalness \citep[p.\,140]{Par89} & .738 & .0020\\ \hline
\end{tabular}\end{center}
\end{table*}

However, for dyads, almost all correlations of the different approaches are
highly statistically significant. Exceptions are complex tonalness
\citep[p.\,140]{Par89} and smoothed relative periodicity, if Pythagorean tuning or
Kirnberger~III are employed, but neither of these tunings is psychophysically
motivated by the just noticeable difference of pitch perception of about 1\%
(cf.~\secref{sec:approx}). In contrast to this, smoothed relative and smoothed logarithmic
periodicity employing one of the rational tunings from \tabref{tab:tun} show
high correlation.

If periodicity detection in the brain is an important mechanism for the
perception of consonance, the periodicity-based method with smoothing should show high
correlation with neurophysiological data as well, and in fact this holds:
\citet{BK09} found auditory nerve and brainstem correlates of musical consonance
and detected that brainstem responses to consonant intervals were more robust and
yielded stronger pitch salience than those to dissonant intervals. They compared
perceptual consonant ratings of $n=9$ musical intervals and estimates of neural
pitch salience derived from the respective brainstem frequency-following
responses. The correlation between the ranking of neural pitch salience
\citep[see][Figure~3]{BK09} and smoothed relative or smoothed logarithmic periodicity (with
rational tuning \#2) is $r=.833$, which is still significant with $p=.0027$.

\opt{long}{Last but not least, it should be noted that the so-called \emph{major profile},
investigated by \citet[Figure~3.1]{Kru90}, also corresponds quite well with the
empirical consonance ranking. Here, an ascending or descending major scale is
presented to test persons. A probe tone comes next. The listener's task is to
rate it as a completion of the scale context. The correlation in this case is
$r=.846$, which is still significant with $p=.0001$.}

\subsection{Triads\opt{long}{ and more}}\label{sec:triad}

\tabref{tab:triad} shows the perceived and computed consonance of
common triads (cf.~\figref{fig:triad}). There are several empirical studies on
the perception of common triads \citep{Rob86,Coo09b,JKL12}. But since the
experiments conducted by \citet[Experiment~1]{JKL12} are the most
comprehensive, because they examined all $55$ possible three-note chords, we
adopt this study as reference for the empirical ranking here. Note that
\citet{JKL12} designed the register of the pitch classes in the chords, i.e. the
spacing, so that the chords spread over about an octave and a half, in order to
make the chords comparable with those that occur in music. All the cited empirical
studies on triads are consistent with the following preference ordering on
triads: major $\prec$ minor $\prec$ suspended $\prec$ diminished $\prec$
augmented, at least for chords in root position. However, the ordinal ratings of
minor and suspended chords do not differ very much. Again, the summary table
(\tabref{tab:cor3}) reveals highest correlations with the empirical ranking for
smoothed relative and smoothed logarithmic periodicity, if the underlying tuning is
psychophysically motivated. Roughness
\citep{HK78,HK79} and the sonance factor \citep{Hof04,Hof08} yield relatively
bad predictions of the perceived consonance of common triads.

\begin{table*}
\caption{Consonance rankings of common triads. The empirical rank is adopted
from \citet[Experiment~1]{JKL12}, where the tones are reduced to one octave in the theoretical analysis here. The
roughness values are taken from \citet[Table~1]{HK79}, where again C4 (middle
C) is taken as the lowest tone. For smoothed relative periodicity and percentage
similarity \citep{GP09}, the frequency ratios from rational tuning \#2 are used. The
dual-process theory \citep[Figure~2]{JKL12} as a cognitive theory only provides
ranks, not numerical raw values.}\label{tab:triad}
\footnotesize
\begin{center}\begin{tabular}{l@{ }r*{5}{c@{\,}c}c} \hline
  & & \multicolumn{2}{c}{Empirical} & & & & & & & Smoothed & Relative & Dual\\
  \multicolumn{2}{l}{Chord class} & \multicolumn{2}{c}{rank} & \multicolumn{2}{c}{Roughness} & \multicolumn{2}{c}{Instability} & \multicolumn{2}{c}{Similarity} & \multicolumn{2}{c}{periodicity} & process\\ \hline
  Major	& \semi{0,4,7}	& 1	& (1.667)	& 3	& (0.1390)	& 1	& (0.624)	& 1-2	& (46.67\%)	& 2	& (4.0)	& 2\\
	  & \semi{0,3,8}	& 5	& (2.889)	& 9	& (0.1873)	& 5	& (0.814)	& 8-9	& (37.78\%)	& 3	& (5.0)	& 1\\
	  & \semi{0,5,9}	& 3	& (2.741)	& 1	& (0.1190)	& 4	& (0.780)	& 5-6	& (45.56\%)	& 1	& (3.0)	& 3\\
  Minor	& \semi{0,3,7}	& 2	& (2.407)	& 4	& (0.1479)	& 2	& (0.744)	& 1-2	& (46.67\%)	& 4	& (10.0)	& 4\\
	  & \semi{0,4,9}	& 10	& (3.593)	& 2	& (0.1254)	& 3	& (0.756)	& 5-6	& (45.56\%)	& 7	& (12.0)	& 5\\
	  & \semi{0,5,8}	& 8	& (3.481)	& 7	& (0.1712)	& 6	& (0.838)	& 8-9	& (37.78\%)	& 10	& (15.0)	& 6\\
  Suspended	& \semi{0,5,7}	& 7	& (3.148)	& 11	& (0.2280)	& 8	& (1.175)	& 3-4	& (46.30\%)	& 5	& (10.7)	& 7\\
	  & \semi{0,2,7}	& 6	& (3.111)	& 13	& (0.2490)	& 11	& (1.219)	& 3-4	& (46.30\%)	& 9	& (14.3)	& 9\\
	  & \semi{0,5,10}& 4	& (2.852)	& 6	& (0.1549)	& 9	& (1.190)	& 7	& (42.96\%)	& 6	& (11.0)	& 8\\
  Diminished	& \semi{0,3,6}	& 12	& (3.889)	& 12	& (0.2303)	& 12	& (1.431)	& 13	& (32.70\%)	& 12	& (17.0)	& 12\\
	  & \semi{0,3,9}	& 9	& (3.519)	& 10	& (0.2024)	& 7	& (1.114)	& 10-11	& (37.14\%)	& 11	& (15.3)	& 10\\
	  & \semi{0,6,9}	& 11	& (3.667)	& 8	& (0.1834)	& 10	& (1.196)	& 10-11	& (37.14\%)	& 8	& (13.3)	& 11\\
  Augmented	& \semi{0,4,8}	& 13	& (5.259)	& 5	& (0.1490)	& 13	& (1.998)	& 12	& (36.67\%)	& 13	& (20.3)	& 13\\ \hline
  \multicolumn{2}{l}{Correlation~$r$} & & & .352 & & .698 & & .802 & & .846 & & .791\\ \hline
\end{tabular}\end{center}
\end{table*}

\begin{table*}
\caption{Correlations of several consonance rankings with empirical ranking for
triads. Only $n=10$ values were available for pure tonalness
\citep[p.\,140]{Par89} and the consonance degree according to
\citet[Figure~6]{ICMPC.Fol12}, because suspended or diminished chords,
respectively, are missing in these references. Thus, we have only $n-2=8$ degrees of freedom in the
calculation of the respective significance values. For some approaches
\citep{Hel63,PL65,KK69,Set05}, the rankings are taken from
\citet[Table~1]{Coo09b}.}\label{tab:cor3}
\footnotesize
\begin{center}
\begin{tabular}{lcc} \hline
  Approach & Correlation $r$ & Significance $p$\\ \hline
  Smoothed relative periodicity (rational tuning \#2) & .846 & .0001\\
  Smoothed logarithmic periodicity (rational tuning \#2) & .831 & .0002\\
  Smoothed logarithmic periodicity (rational tuning \#1) & .813 & .0004\\
  Smoothed relative periodicity (rational tuning \#1) & .808 & .0004\\
  Percentage similarity \citep{GP09} & .802 & .0005\\
  Dual process \citep[Figure~2]{JKL12} & .791 & .0006\\
  Consonance value \citep{Bre05} & .755 & .0014\\
  Consonance degree \citep[Figure~6]{ICMPC.Fol12} & .826 & .0016\\
  Dissonance curve \citep{Set05} & .723 & .0026\\
  Instability \citep[Table~A2]{CF06} & .698 & .0040\\
  Smoothed gradus suavitatis \citep{Eul39} & .690 & .0045\\
  Sensory dissonance \citep{KK69} & .607 & .0139\\
  Tension \citep[Table~A2]{CF06} & .599 & .0153\\
  Pure tonalness \citep[p.\,140]{Par89} & .675 & .0162\\
  Critical bandwidth \citep{PL65} & .570 & .0210\\
  Temporal dissonance \citep{Hel63} & .503 & .0399\\
\opt{long}{Relative prevalence of chord types \citep[p.~421]{Ebe94} & .481 & .0482\\}
  Sonance factor \citep{Hof04,Hof08} & .434 & .0692\\
  Roughness \citep[Table~1]{HK79} & .352 & .1193\\ \hline
\end{tabular}\end{center}
\end{table*}

The data sets in \citet{JKL12} suggest further investigations. So
\tabref{tab:complete} shows the analysis of all possible three-tone chords in
root position \citep[Figure~2]{JKL12}. As one can see, the correlation between
the empirical rating and the predictions of the dual-process theory is very
high ($r=.916$). This also holds for smoothed logarithmic periodicity but not that much
for smoothed relative periodicity, in particular, if the correlation between the ordinal
rating and the concrete smoothed periodicity values are taken, which are shown in
parentheses in \tabref{tab:complete} ($r=.810$ versus $r=.548$). This justifies our
preference for smoothed logarithmic as opposed to smoothed relative periodicity, because the former notion is
motivated more by neuroacoustical results, namely that the spatial structure of
the periodicity-pitch representation in the brain is organised as a logarithmic
periodicity map. \opt{long}{\citet[Experiment~2]{JKL12} also provide data for
$n=48$ four-tone chords. A summary analysis is shown in \tabref{tab:cor4},
indicating once again high correlation for several measures of harmoniousness,
including logarithmic periodicity.}

\begin{table*}
\caption{Consonance correlation for the complete list of triads in root position
with spacing of chords as given by \citet[Figure~2, $n=19$]{JKL12}. The
ordinal rating and the numerical values
of the considered measures are given in parentheses. The correlation and
significance values written in parentheses refer to the ordinal rating, while
the ones outside parentheses just compare the respective rankings. The roughness
values are taken from \citet[Figure~2]{JKL12}\opt{long}{, who employ the implementation
available at \protect\url{http://www.uni-graz.at/richard.parncutt/computerprograms.html},
based on the research reported in \citet{BPL96}}. In all other cases, rational tuning \#2
is used as the underlying tuning for the respective frequency ratios.}\label{tab:complete}
\footnotesize
\begin{center}
\begin{tabular}{l@{ }r*{5}{@{ }c@{\,}c}@{ }c@{ }} \hline
  & & \multicolumn{2}{c}{Empirical} & & & & & \multicolumn{2}{c}{Smoothed relative} & Smoothed & logarithmic & Dual\\
  Chord\,\# & Semitones & \multicolumn{2}{c}{rank} & \multicolumn{2}{c}{Roughness} & \multicolumn{2}{c}{Similarity} & \multicolumn{2}{c}{periodicity} & \multicolumn{2}{c}{periodicity} & process\\ \hline
  1a (major) & \semi{0,16,19}	& 1	& (1.667)	& 2	& (0.0727)	& 1	& (64.44\%)	& 1	& (2.0)		& 1	& (1.000)	& 1\\
  2a	& \semi{0,19,22}	& 3	& (2.481)	& 1	& (0.0400)	& 5	& (52.59\%)	& 8-9	& (12.3)	& 6-7	& (3.133)	& 2\\
  3a	& \semi{0,16,22}	& 4-5	& (2.630)	& 3	& (0.0760)	& 9	& (38.62\%)	& 15-16	& (19.0)	& 8	& (3.271)	& 3\\
  4a	& \semi{0,15,22}	& 6	& (2.926)	& 4	& (0.0894)	& 8	& (39.26\%)	& 8-9	& (12.3)	& 6-7	& (3.133)	& 4\\
  5a (minor) & \semi{0,15,19}	& 2	& (2.407)	& 5	& (0.0972)	& 3-4	& (55.56\%)	& 4	& (5.0)		& 4	& (2.322)	& 5\\
  6a (suspended) & \semi{0,7,14}	& 7	& (3.148)	& 6	& (0.0983)	& 3-4	& (55.56\%)	& 6-7	& (11.7)	& 5	& (2.918)	& 6\\
  7a	& \semi{0,19,23}	& 8	& (3.370)	& 7	& (0.1060)	& 2	& (56.67\%)	& 2-3	& (4.0)		& 2-3	& (2.000)	& 7\\
  8a	& \semi{0,16,23}	& 4-5	& (2.630)	& 8	& (0.1097)	& 6	& (52.22\%)	& 2-3	& (4.0)		& 2-3	& (2.000)	& 8\\
  9a (diminished) & \semi{0,15,18}	& 11	& (3.889)	& 16	& (0.2214)	& 17	& (28.57\%)	& 13	& (17.0)	& 14	& (3.786)	& 9\\
  10a	& \semi{0,11,14}	& 13	& (3.963)	& 9	& (0.1390)	& 18	& (28.33\%)	& 10-11	& (14.3)	& 12	& (3.585)	& 10\\
  11a	& \semi{0,18,22}	& 15	& (5.148)	& 12	& (0.1746)	& 14	& (30.05\%)	& 15-16	& (19.0)	& 11	& (3.540)	& 11\\
  12a	& \semi{0,17,23}	& 12	& (3.926)	& 13	& (0.1867)	& 12	& (34.37\%)	& 6-7	& (11.7)	& 10	& (3.497)	& 12\\
  13a	& \semi{0,14,17}	& 9	& (3.481)	& 14	& (0.1902)	& 11	& (36.11\%)	& 5	& (10.0)	& 9	& (3.308)	& 13\\
  14a	& \semi{0,15,26}	& 10	& (3.630)	& 17	& (0.2485)	& 13	& (33.52\%)	& 12	& (15.7)	& 15	& (3.800)	& 14\\
  15a	& \semi{0,11,18}	& 14	& (4.815)	& 18	& (0.2639)	& 10	& (36.90\%)	& 17-18	& (25.7)	& 17	& (4.571)	& 15\\
  16 (augmented) & \semi{0,16,20}	& 16	& (5.259)	& 10	& (0.1607)	& 7	& (41.67\%)	& 10-11	& (14.3)	& 13	& (3.655)	& 16\\
  17a	& \semi{0,15,23}	& 17-18	& (5.593)	& 11	& (0.1727)	& 16	& (28.89\%)	& 17-18	& (25.7)	& 18	& (4.655)	& 17\\
  18a	& \semi{0,20,23}	& 19	& (5.630)	& 15	& (0.2164)	& 15	& (29.44\%)	& 14	& (17.7)	& 16	& (3.989)	& 18\\
  19a	& \semi{0,14,25}	& 17-18	& (5.593)	& 19	& (0.3042)	& 19	& (19.93\%)	& 19	& (101.0)	& 19	& (5.964)	& 19\\ \hline
  \multicolumn{2}{l}{Correlation~$r$} & & & .761 & (.746) & .760 & (.765) & .713 & (.548) & .867 & (.810) & .916\\
  \multicolumn{2}{l}{Significance~$p$} & & & .0001 & (.0001) & .0001 & (.0001) & .0003 & (.0075) & .0000 & (.0000) & .0000\\ \hline
\end{tabular}
\end{center}
\end{table*}

\opt{long}{\begin{table*}
\caption{Correlations of several consonance rankings with ordinal values of
empirical rating for $n=48$ selected four-note chords, spread over more than one
octave \citep[Figure~3]{JKL12}. For the $\Omega$ measure \citep{ICMPC.Sto12b},
percentage similarity \citep{GP09}, and gradus suavitatis \citep{Eul39}, once
again the frequency ratios from rational tuning \#2 are used.}\label{tab:cor4}
\footnotesize
\begin{center}
\begin{tabular}{lcc} \hline
  Approach & Correlation $r$ & Significance $p$\\ \hline
  Dual process \citep[Figure~3]{JKL12} & .895 & .0000\\
  Smoothed $\Omega$ measure \citep{ICMPC.Sto12b} & .824 & .0000\\
  Smoothed gradus suavitatis \citep{Eul39} & .785 & .0000\\
  Smoothed logarithmic periodicity (rational tuning \#2) & .758 & .0000\\
  Smoothed logarithmic periodicity (rational tuning \#1) & .754 & .0000\\
  Percentage similarity \citep{GP09} & .734 & .0000\\
  Smoothed relative periodicity (rational tuning \#2) & .567 & .0000\\
  Smoothed relative periodicity (rational tuning \#1) & .531 & .0001 \\
  Roughness \citep{HK78,HK79} & .402 & .0023\\ \hline
\end{tabular}\end{center}
\end{table*}}

\subsection{From chords to scales}

In contrast to many other approaches, the periodicity-based method with smoothing can easily
be applied to scales and yields meaningful results \citep{ICMPC.Sto09}.

\begin{figure*}[t]
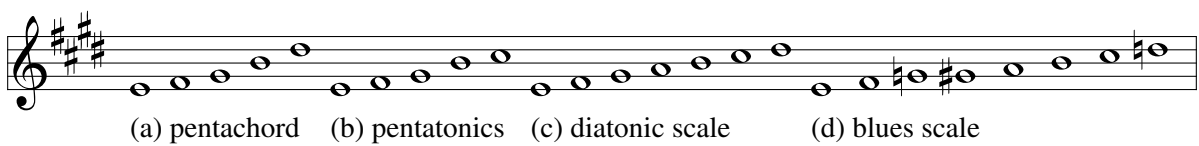

\centering
  \begin{music}
    \instrumentnumber{1}\setsign{1}{4}\nobarnumbers
    \startextract
	\notesp\zsong{(a) pentachord}\wh{efg'bd}\enotes
        \notesp\zsong{(b) pentatonics}\wh{efg'bc}\enotes
        \notesp\zsong{(c) diatonic scale}\wh{efg'abcd}\enotes
        \Notes\zsong{(d) blues scale}\wh{ef=g^g'abc=d}\enotes
    \endextract
  \end{music}
\caption{Harmonies (scales) with more than three tones.}\label{fig:scale}
\end{figure*}

\begin{exmp}
\figref{fig:scale}(a) shows the pentachord Emaj$7/9$ (with E4 as the lowest tone),
classically built from a stack of thirds, standard in jazz music. It is the
highest ranked harmony with $5$ out of $12$ tones ($\overline{\log_2(h)} \approx 4.234$ with
respect to rational tuning \#1, $\overline{\log_2(h)} \approx 3.751$ with respect to rational
tuning \#2). Although in most cases this pentachord is most likely not heard in
a bitonal manner, it may be alternatively understood as the superposition of the
major triads E and B, which are in a tonic-dominant relationship according to
classical harmony theory. As just said, it appears in the front rank of all
harmonies consisting of $5$ out of $12$ tones, which does not hold for a
superposition of a random chord sequence. Thus, also for chord progressions the
periodicity-based method with smoothing yields meaningful results. Nevertheless, this point has
to be investigated further. All harmonies shown in \figref{fig:scale} have low,
i.e. good smoothed periodicity values, ranking among the top 5\% in their tone
multiplicity category with respect to rational tuning \#1. This holds for the
pentatonics ($5$ tones, $\overline{\log_2(h)} \approx 5.302$), the diatonic scale ($7$ tones,
$\overline{\log_2(h)} \approx 6.453$), as well as the blues scale ($8$ tones, $\overline{\log_2(h)} \approx 7.600$).
\end{exmp}

\citet{TT13} investigated the perceived consonance of diatonic scales.
\tabref{tab:church} lists all classical church modes, i.e. the diatonic scale
and its inversions. The cognitive model of the perception of diatonic scales
introduced by \citet{TT13} results in a 100\% correlation with the empirical
data. Although the correlation for smoothed logarithmic periodicity obviously is not that
good, it still shows high correlation. Nonetheless, the major scale (Ionian,
cf.~\figref{fig:scale}(c)) appears in the front rank of $462$ possible scales with
$7$ out of $12$ tones with respect to smoothed relative and smoothed logarithmic periodicity. In
addition, in contrast to more cognitive theories on harmony perception, the
periodicity-based method with smoothing introduced in this article does not presuppose any principles
of tonal music, e.g. the existence of diatonic scales or the common use of the
major triad. They may be derived from underlying, more primitive mechanisms,
namely periodicity detection in the human (as well as animal) brain. 

Interestingly, with rational tuning \#1 as basis, the results for scales
(\tabref{tab:church}) are better for smoothed logarithmic periodicity than with rational
tuning \#2 ($r=.964$ versus $r=.786$). In the former case, the seven classical church
modes even appear in the very front ranks of their tone multiplicity category.
The correlation of percentage similarity is low. But despite this, the
diatonic scale and its inversions are among the $50$ heptatonic scales whose
intervals conform most closely to a harmonic series out of $4 \cdot 10^7$
examined possibilities \citet[Table~3]{GP09}, which is of course a meaningful
result.

\begin{table*}
\caption{Rankings of common heptatonic scales (church modes), i.e. with $7$ out
of $12$ tones. As empirical rating, the overall preference for the classical
church modes is adopted \citep[Figure~10]{TT13}. For the sonance factor
\citep{Hof04,Hof08}, again C4 (middle C) is taken as the lowest tone. For
percentage similarity, the values are taken directly from \citet[Table~3]{GP09}.}\label{tab:church}
\footnotesize
\begin{center}
\begin{tabular}{*{6}{c@{\,}c}} \hline
  & & & & & & & & \multicolumn{2}{c}{Smoothed} & \multicolumn{2}{c}{Smoothed}\\
  Mode & Semitones & \multicolumn{2}{c}{Empirical} & \multicolumn{2}{c}{Sonance} & \multicolumn{2}{c}{Similarity} & \multicolumn{2}{c}{log. periodicity} & \multicolumn{2}{c}{log. periodicity}\\
 & & \multicolumn{2}{c}{rank} & \multicolumn{2}{c}{factor} & & & \multicolumn{2}{c}{(Rational tuning \#1)} & \multicolumn{2}{c}{(Rational tuning \#2)}\\ \hline
  Ionian	& \semi{0,2,4,5,7,9,11}	& 1	& (0.83)	& 4	& (0.147)	& 3	& (39.61\%)	& 1	& (6.453)& 1	& (5.701)	\\
  Mixolydian	& \semi{0,2,4,5,7,9,10}	& 2	& (0.64)	& 1.5	& (0.162)	& 6	& (38.59\%)	& 3	& (6.607)& 4	& (5.998)	\\
  Lydian	& \semi{0,2,4,6,7,9,11}	& 3	& (0.58)	& 1.5	& (0.162)	& 5	& (38.95\%)	& 2	& (6.584)& 2	& (5.830)	\\
  Dorian	& \semi{0,2,3,5,7,9,10}	& 4	& (0.40)	& 3	& (0.152)	& 2	& (39.99\%)	& 4	& (6.615)& 3	& (5.863)	\\
  Aeolian	& \semi{0,2,3,5,7,8,10}	& 5	& (0.34)	& 6	& (0.138)	& 4	& (39.34\%)	& 5	& (6.767)& 7	& (6.158)	\\
  Phrygian	& \semi{0,1,3,5,7,8,10}	& 6	& (0.21)	& 7	& (0.126)	& 1	& (40.39\%)	& 6	& (6.778)& 5	& (6.023)	\\
  Locrian	& \semi{0,1,3,5,6,8,10}	& 7	& 		& 5	& (0.142)	& 7	& (37.68\%)	& 7	& (6.790)& 6	& (6.033)	\\ \hline
  \multicolumn{2}{l}{Correlation~$r$} & & & .667 & & .036 & & .964 & & .786 & \\
  \multicolumn{2}{l}{Significance~$p$} & & & .0510 & & .4697 & & .0002 & & .0181 & \\ \hline
\end{tabular}
\end{center}
\end{table*}

\section{Discussion and conclusions}\label{sec:sum}

\subsection{Summary}

We have seen in this article that harmony perception can be explained well by 
considering the periodic structure of harmonic sounds, which can be computed from
the frequency ratios of the intervals in the given harmony. The
results presented show the highest correlation with empirical results and thus contribute to
the discussion of the consonance/dissonance of musical chords and scales. We
conclude that there is a strong neuroacoustical and psychophysical basis for
harmony perception including chords and scales.

\subsection{Limitations of the approach}

The periodicity-based method with smoothing, as presented in this article, clearly has
limitations. First, the available empirical studies on harmony perception used
in this article \citep{SHP03,JKL12,TT13} in general take the average over all
participants' ratings. Thus, individual differences in perception are
neglected, e.g. the influence of culture, familiarity, or musical training.
However, some studies report that especially the number of years of musical training has
a significant effect on harmony perception, although periodicity detection
remains an important factor that is used to a different extent by musicians and
non-musicians \citep{HH84,LS+09,MM+13}.

Second, we do not consider in detail the context of a harmony in a musical piece
during the periodicity-based analyses with smoothing in this article. Therefore, \citet{PH11}
attempt to explain the perception of musical harmonies as a
holistic phenomenon, covering a broad spectrum, including the conception of
consonance/dissonance as pleasant/unpleasant, and the history in Western music
and music theory, emphasising that consonance/dissonance should be discussed
along several dimensions such as tense/relaxed, familiar/unfamiliar, and
tonal/atonal. Nonetheless, periodicity may be used as one constituent in
explaining harmony perception, even with respect to the historical development
of (Western) music, by assuming different levels of harmonic complexity, changing
over time \citep[pp.\,57-58]{Par89}.

Third, in this article we adopt the equal temperament as a reference system for
tunings, which is the basis of Western music. However, non-Western scales, e.g.
in Turkish classical music with Makam melody types or tone scales of recordings
of traditional Central African music \citep[see
\url{http://music.africamuseum.be/} and][]{ICMPC.MCL09}, do not seem to be
based on equal temperament tunings. Nevertheless, they can be analysed by the
periodicity-based method with smoothing, predicting also
relatively good, i.e. low values of consonance for these scales \citep{Sto10d}.
\opt{long}{\citet{Lap17} also studied music from different cultures and epoques,
including Turkish, Arabic, and Chinese music. He concludes that for microtonal
scales (smoothed logarithmic) periodicity as harmoniousness measure should be
restricted to smaller groups of tones, which he calls \emph{words}, e.g. three
consecutive tones. This point clearly must be further investigated. The
approaches by \citet{GP09} and \citet{HB11} are also applicable to non-Western
music, too.

Relative and logarithmic periodicity of harmonies can be computed without reference
to any tuning by approximating frequency ratios within a band of $d = \pm$1\%
employing the procedure from \secref{sec:approx} (see \defref{def:extended}
below). A re-implementation of the Prolog code (cf. \secref{sec:eval}) for this
generalized periodicity measure in GNU~Octave \citep{EB+17}, which is largely
compatible with the scientific programming language Matlab \citep{HH17}, is
available (also) at \url{http://artint.hs-harz.de/fstolzenburg/harmony/}.

\begin{defn}\label{def:extended}
Let $H = \{ f_1, \dots, f_n \}$ be a harmony and $H_k = \{ f_1/f_k, \dots,
f_n/f_k \}$ for $1 \le k \le n$. Then a harmoniousness measure $\H$ can be
\emph{smoothed} for general harmonies, i.e. without reference to a tuning
system or semitone sets (in contrast to \defref{def:smooth}), as follows:
	\[ \widetilde{\H}(H) = \frac{1}{n} \sum_{k=1}^n \H(H_k) \]
If $\H$ is relative or logarithmic periodicity, then each (real) frequency ratio
$F$ in a harmony class $H_k$ has to be made rational, and octave equivalence can
be taken into account. For this, let $F = 2^m \cdot F_1$ for a (possibly
negative) integer $m$ such that $1 \le F_1 \le 2$, and $F_2$ be the rational
approximation of $F_1$ (according to \secref{sec:approx}) with some given
maximal deviation $d$. Then the integer ratio $F_3 = 2^m \cdot F_2$ (in lowest
terms) shall be used instead of $F$ in the computation.
\end{defn}}

\subsection{Future work}

Future work should concentrate on even more exhaustive empirical experiments on
harmony perception, in order to improve the significance of and confidence in the
statistical analyses, including more detailed investigations of chord
progressions, possibly employing different tunings and timbres, and taking into
account the historical development of Western music and beyond. Last but not
least, the working of the brain with respect to auditory processing still must
be better understood \citep[see e.g.][]{PU+02}. In consequence, models of the
brain that take temporal properties into account should be investigated, as claimed
also by \citet{Roe08}.\opt{long}{ For this, (artificial) neural networks that have
this property \citep{Car01a,BL06,Hay08,SR_09a,VL+04} could be considered
further.

\pagebreak}

\phantomsection\pdfbookmark[1]{Acknowledgements}{toc}
\section*{Acknowledgements}

This article is a completely revised and extended version of previous work
\citep{ICMPC.Sto09,Sto10d,ICMPC.Sto12b,Sto14b}. I would like to thank Wolfgang
Bibel, Peter A. Cariani, Norman D. Cook, Martin Ebeling, Thomas Fiore, Adrian
Foltyn, Sergey Gaponenko, Ludger J. Hofmann-Engl, Phil N. Johnson-Laird, Gerald
Langner, Andrew J. Milne, Florian Ruh, Tilla Schade, and Marek \v{Z}abka,
as well as several anonymous referees for helpful discussions, hints, and
comments on this article or earlier versions thereof.

\bibliographystyle{apalike-url}
\phantomsection\pdfbookmark[1]{References}{toc}
\bibliography{music,frieder,stolzen}

\end{document}